\documentclass[aps,prd,nofootinbib,unsortedaddress,showpacs,twocolumn]{revtex4}
\usepackage{graphicx}

\newcommand{\be}{\begin{equation}}
\newcommand{\ee}{\end{equation}}
\newcommand{\bea}{\begin{eqnarray}}
\newcommand{\eea}{\end{eqnarray}}
\newcommand{\simge}{\gtrsim}
\newcommand{\simle}{\lesssim}

\newcommand{\GeV}{\mbox{GeV}}

\newcommand{\lsim}{\raisebox{-0.13cm}{~\shortstack{$<$ \\[-0.07cm] $\sim$}}~}
\newcommand{\gsim}{\raisebox{-0.13cm}{~\shortstack{$>$ \\[-0.07cm] $\sim$}}~}

\begin{document}
\begin{flushright}
{\small \mbox{PM/04-20}}
\end{flushright}

\title{Gravitino dark matter in gauge mediated supersymmetry breaking}

\author{{\sf Karsten Jedamzik} ${}^{a}$, {\sf Martin Lemoine} ${}^{b}$, 
{\sf Gilbert Moultaka} ${}^{a}$}
\affiliation{~~}

\author{
{\small {\it ${}^{a}$ Laboratoire de Physique Th\'eorique \& Astroparticules, \\ CNRS
UMR 5207, Universit\'e Montpellier II, \\
Place E. Bataillon, F-34095 Montpellier Cedex 5,
France} }}
\affiliation{~~}

\author{
{\small {\it ${}^{b}$ Institut d'Astrophysique de Paris,\\  UMR 7095 CNRS, 
Universit\'e Pierre \& Marie Curie, \\
98 bis boulevard Arago, F-75014 Paris, France}} }

\begin{abstract}
This paper investigates the parameter space of theories with gauge
mediated supersymmetry breaking leading to gravitino (cold) dark
matter with mass $m_{3/2}\sim 1\,{\rm keV}\rightarrow10\,{\rm
MeV}$. We pay particular attention to the cosmological r\^ole of
messenger fields. Cosmology requires that these messengers decay to
the visible sector if the lightest messenger mass $M_X\gtrsim
30\,$TeV. We then examine the various possible messenger number
violating interactions allowed by the symmetries of the theory and by
phenomenology. Late messenger decay generally results in entropy
production hence in the dilution of pre-existing gravitinos. We find
that in $SU(5)$ grand unification only specific messenger-matter
couplings allow to produce the required amount of gravitino dark
matter particles. Gravitino dark matter with the correct abundance is
however expected in larger gauge groups such as $SO(10)$ for generic
non-renormalizable messenger-matter interactions and for arbritrarily
high post-inflationary reheating temperatures.
\end{abstract}


\maketitle

\setcounter{secnumdepth}{3}

\section{Introduction}

Supersymmetric extensions of the standard electroweak model of
particle physics come with a variety of appealing byproducts, such as
the stabilization of the Higgs mass against radiative corrections,
radiatively induced electro-weak symmetry breaking at the electro-weak
scale, and the possibility of $SU(3)_c\times SU(2)_L\times U(1)_Y$
gauge coupling unification at a sufficiently high energy scale.  Most
of these features occur rather naturally after supersymmetry (SUSY)
has been broken softly, and it is believed that ultimately such a
breaking has to occur spontaneously within some theory describing all
four fundamental interactions.

Despite the lack of a particularly compelling model, there is a number
of proposals with various theoretical and phenomenological merits,
where the spontaneous SUSY breaking usually takes place dynamically in
a (hidden) sector of the theory which does not contain the standard
model particles. Models can be classified according to the origin of
SUSY breaking and of the soft terms, i.e. how the breaking of
supersymmetry is transmitted to the low energy (visible) sector. In
the so-called gauge mediated supersymmetry breaking (GMSB) models,
this transmission is induced by renormalizable gauge
interactions~\cite{GMSB0, GMSB} (see~\cite{GR99} for a review). Particularly
attractive features of these scenarios are the natural suppression of
neutral current flavor changing interactions as well as a highly
predictive mass spectrum that will be put to test in forthcoming
collider experiments.

 In theories with gauge mediation, the lightest supersymmetric
particle (LSP) is the gravitino and its mass can lie anywhere in the
range $m_{3/2}\sim 1\,{\rm eV}\rightarrow 1\,{\rm GeV}$. Such a light
gravitino (i.e. when $m_{3/2}\simge 1\,{\rm keV}$) is traditionally
associated with a cosmological catastrophe. Indeed many studies have
examined the production of light gravitinos in the early Universe in
order to place stringent bounds on the post-inflationary reheating
temperature (see~\cite{TMY93,M95,CHKL99} and references therein).  Few
studies have contemplated the possibility that this gravitino LSP
could make up the dark matter of the
Universe~\cite{BM03,FY02,FY02b,FIY04, IY04, AD05}, mainly because in
the most naive model one needs to adjust the reheating temperature as
a function of the gravitino mass in order to obtain the required dark
matter relic density. However Refs.~\cite{BM03,FY02} have recognized
the important cosmological r\^ole of the messenger particles that are
part of the spectrum of all GMSB theories. In particular, it has been
shown that the late decay of the lightest messenger to visible sector
particles can induce a substantial amount of entropy production which
would result in the dilution of the predicted gravitino abundance. As
a result, the light gravitino problem could be turned into a light
gravitino blessing, i.e. one would obtain suitable gravitino dark
matter for arbitrarily high reheating temperatures.

  These studies~\cite{BM03,FY02} have focused on two specific
couplings between the messenger and visible sectors. This motivates us
to examine in more generality the possibility of producing the right
amount of gravitino dark matter in GMSB scenarios. We do so in the
present paper by considering all messenger-matter interactions allowed
by the gauge symmetries of the theory and by phenomenology and by
considering their impact on the gravitino abundance. The present study
thus aims at being more exhaustive than prior investigations; on the
way we will also improve on some results previously obtained. In
particular we show that the coupling introduced in Ref.~\cite{BM03}
does not appear in minimal GMSB models and that multi-goldstino
production channels modify substantially the results of
Ref.~\cite{FY02}. Finally we also take into account stringent
constraints from big-bang nucleosynthesis and large scale structure
formation.

  Our study is similar in spirit to those conducted for neutralino
dark matter in minimal supergravity~\cite{neutralino}. However since
the present paper is of an exploratory nature, we approximate the mass
spectrum of GMSB models by three parameters: the messenger mass scale
$M_X$, the supersymmetric particles mass scale $M_{\rm SUSY}\sim
1\,$TeV and the gravitino mass $m_{3/2}$. In particular, we treat
$m_{3/2}$ and $M_X$ as the fundamental parameters in our search for
gravitino dark matter. We also discuss the influence of the nature and
mass of the next-to-lightest supersymmetric particle
(NLSP). Furthermore we calculate the velocity of the dark matter
gravitinos in order to examine whether this dark matter is hot, warm
or cold. For simplicity we assume that $R-$parity holds. However, we
will show that, when $m_{3/2}\lesssim 10\,{\rm MeV}$, our results
remain valid even if $R-$parity is violated. In this range the
gravitino lifetime becomes much longer than the age of the Universe,
so that it can be considered as stable on our cosmological time and
constraints from diffuse background distortions are
eluded~\cite{TY00,MC02}. Finally we note that Ref.~\cite{EOSS03,FST04} has
discussed the possibility of gravitino dark matter in minimal
supergravity models (with gravity mediation of supersymmetry
breaking). The cosmology of these models is different from that of
GMSB scenarios as the gravitino is heavier ($m_{3/2}\gtrsim 10\,$GeV)
and there are no messenger particles. It is found in these studies
that only a limited region of parameter space can satisfy the big-bang
nucleosynthesis constraints and that the reheating temperature must be
tuned in order to obtain the required dark matter relic density.

  The plan of this paper is as follows.  In Section~\ref{sec:2} we
review the basics of gauge mediation models and discuss the nature of
the lightest messenger particle which plays a crucial r\^ole in our
analysis.  In Section~\ref{sec:3} we discuss the numerous channels of
light gravitino production in the early Universe as well as gravitino
dilution due to late decay of a massive particle and the relevant
cosmological constraints. In Section~\ref{sec:4} we survey the various
renormalizable and non-renormalizable messenger number violating
interactions allowed by the gauge symmetry of the theory and discuss
their consequences with respect to the light gravitino problem and
gravitino dark matter. Finally in Section~\ref{sec:5}, we discuss
various perspectives, notably with respect to the case of $SO(10)$
grand unification. We restrict ourselves to GMSB scenarios in the
framework of $N=1$ $D=4$ supergravity and use natural units
$\hbar=c=k_{\rm B}=1$; $m_{\rm Pl}\simeq 2.4\times 10^{18}\,$GeV
denotes the reduced Planck mass.

\section{Messenger sector}\label{sec:2}

\subsection{Gauge mediation of SUSY breaking}
\noindent
Gauge-mediated SUSY breaking (GMSB) \cite{GMSB,GR99}, is usually
implemented by adding a term \be W = S\Phi_M\overline{\Phi}_M + \Delta
W(S,Z_i) \label{eq:W1} \ee to the superpotential where $\Phi_M$ and
$\overline{\Phi}_M$ are messenger left chiral superfields with
$SU(3)\times SU(2)\times U(1)$ quantum numbers, whereas the spurion
left chiral superfield $S$ and the secluded sector $Z_i$ fields are
electroweak- and strong- interactions singlets. Upon the development
of a non-vanishing {\it vev} $\langle S \rangle$ of the scalar
component of the spurion superfield and a SUSY-breaking expectation
value of the spurion auxiliary field $F_S$, due to unspecified
dynamics in the secluded sector $\Delta W(S,Z)$, fermionic messengers
combine into Dirac fermions of mass $M_{X,1/2}= M_X \equiv \langle
S\rangle$, whereas their bosonic partners mix in a mass matrix of
eigenstates $\phi$ and $\overline\phi$ with masses $M_\phi =
M_X(1-F_S/M_X^2)^{1/2}$ and
$M_{\overline\phi}=M_X(1+F_S/M_X^2)^{1/2}$. In terms of the messengers
bosonic components, $\phi=(-\Phi^*_M + \overline\Phi_M)/\sqrt{2}$ and
$\overline\phi=(\Phi_M+\overline\Phi^*_M)/\sqrt{2}$. Note that $\phi$
denotes a set of scalar fields transforming under some representation
of the grand unified gauge group, and that the mass degeneracy of this
multiplet is lifted by $D-$terms and radiative corrections.

Since the messengers share the standard model gauge interactions, the
gaugino and scalar spartners acquire mass at the one- and two-loop
levels respectively:
\begin{equation}
\tilde m_{1/2}\sim \left({\alpha\over 4\pi}\right){F_S\over
M_X}\,,\quad\quad \tilde m_0^2 \sim \left({\alpha\over 4\pi}\right)^2
\left({F_S\over M_X}\right)^2,
\label{eq:GMSBmass}
\end{equation}
hence the quantity $\Lambda = F_S/M_X$ is the supersymmetry breaking
scale in the visible sector.  Provided $F_S/M_X \approx 100\,$TeV,
this generates the required order of magnitude for the soft
parameters\footnote{In the present exploratory study, we do not
address the detailed features of the GMSB mass spectrum and the
related electroweak symmetry breaking and fine-tuning issues.}. Note
that $F_S < M_X^2$ is mandatory otherwise one of the messengers bosons
acquires a negative mass squared.  This also implies $M_X\gtrsim
100\,$TeV, and for $M_X\gg 100\,$TeV, $F_S \ll M_X^2$, hence $M_X$
sets the mass scale for the messenger sector. In particular, $M_\phi
\approx M_X$, and in the following no distinction will be made between
these two mass scales, except where otherwise noted.

The gravitino mass is related to the fundamental SUSY breaking scale,
$m_{3/2}\equiv F/\sqrt{3}m_{\rm Pl}$, with $F\equiv F_S + \sum_i
F_{Z_i}$ the sum of $F-$terms in the secluded sector.  We define the
parameter $k\equiv F_S/F \leq 1$, so that
$m_{3/2}=F_S/(k\sqrt{3}m_{\rm Pl})$. In direct gauge mediated
scenarios, one expects $k\lesssim 1$, whereas in scenarios in which
the transmission of supersymmetry breaking to the messenger sector is
loop suppressed one may find $k\ll 1$. Note that one can also relate
the parameters $k$, $M_X$ and $m_{3/2}$ via the following formula:
$m_{3/2} = \Lambda M_X/\left (k\sqrt{3}m_{\rm Pl}\right)$. Since
$\Lambda$ is tied to the electroweak scale, the latter equation allows
to eliminate one parameter, which we choose to be $k$, in terms of
$m_{3/2}$ and $M_X$, which we will treat as the fundamental
parameters.

\subsection{Lightest messenger}

Taken at face value, GMSB scenarios generically lead to a cosmological
catastrophe, as they predict that the lightest messenger should
overclose the Universe\footnote{Obviously, similar problems can in
principle arise also for the $Z_i$ fields present in $\Delta W$ if the
lightest $Z_i$ mass is larger than $\sim
100\,$TeV~\cite{DGP96}. However, the issue becomes much more
model-dependent here, and we will thus assume for simplicity that
secluded sector fields can decay rapidly to the spurion field $S$. The
latter field is free from such cosmological problems: even though it
can be either heavier or lighter than the messengers [see the
discussion following Eq.(3)], in the first case it decays at
tree-level to messengers, while in the second it decays to gauge
bosons and gauginos fairly quickly through one-loop effects which are
not suppressed by supersymmetry.}. In effect, messenger gauge
interactions as well as those derived from Eq.~(\ref{eq:W1}) conserve
messenger number so that the lightest messenger (a boson) is stable in
this minimal version of the theory.  As messengers can be easily
produced in the primordial plasma thanks to their gauge interactions,
their present day abundance is given by the result of a thermal
freeze-out of messenger annihilation (akin to the well-known
neutralino LSP freeze-out in gravity mediated SUSY breaking).

Through explicit computation, one can show that the lightest messenger
generically overcloses the Universe unless its mass $M_X\lesssim
10^4\,$GeV~\cite{DGP96}.  By lightest messenger, it is 
understood the lightest component of $\phi$ after taking into account
$D-$terms and radiative corrections. Henceforth we denote this
component by $X$. It has been shown that if the messengers sit in
$\mathbf{5}+\mathbf{\overline{5}}$ representations\footnote{Messengers
sitting in complete GUT representations  preserve automatically 
 gauge coupling unification.} of $SU(5)$, the lightest
messenger carries the gauge charges of a sneutrino $\tilde\nu_L$, and
its relic abundance would be of the right order of magnitude provided
its mass $\sim 10-30\,$TeV~\cite{DGP96,HMR04}. Note that the mass
scale in the messenger sector is {\it a priori} unconstrained, since 
phenomenology constrains the ratio $F_S/M_X$ as discussed previously.
If the
messengers sit in $\mathbf{10}+\mathbf{\overline{10}}$ representations
of $SU(5)$, the lightest messenger carries the gauge charges of a
selectron $\tilde e_R$.  Charged dark matter is forbidden by
cosmology~\cite{CDM} and moreover this messenger would overclose the
Universe for typical values of $M_X$. Finally, if the messengers sit
in $\mathbf{16}+\mathbf{\overline{16}}$ representations of $SO(10)$,
the lightest messenger is a $\tilde\nu_R-$like $SU(3)\times
SU(2)\times U(1)$ singlet. The next-to-lightest messenger can only
decay to the lightest messenger by GUT scale suppressed interactions,
and its lifetime of order $10^{10}\,{\rm yrs}\,(M_X/100{\rm
TeV})^{-5}$ is so long that its decay produces unacceptable
distortions of the diffuse backgrounds~\cite{DGP96}.

  Cosmology thus forbids the lightest messenger to be stable unless
messengers can be diluted to a very low abundance or the lightest
messenger happens to have mass $M_X\sim 10-30\,$TeV. If the
post-inflationary reheating temperature is larger than $M_X$ and no
late-time entropy production occurs, then messenger number must be
violated, i.e. the Lagrangian of the theory must contain additional
messenger-matter interactions. However such terms can spoil the
phenomenological successes of the minimal model, in particular the
absence of flavor changing neutral currents or an adequate pattern of
electroweak symmetry breaking. One is thus tempted to believe that
such further messenger interactions with visible sector particles are
rather weak, possibly resulting from non-renormalizable
operators. This will be discussed in more detail below.

Delayed messenger decay can have dramatic consequences for the
gravitino problem and/or the possibility of gravitino dark matter. If
a non-relativistic species comes to dominate the energy density of the
early Universe and subsequently decays into visible sector particles,
a secondary epoch of reheating results and is concommitant with the
dilution of any pre-existing relics, such as gravitinos.  The
abundance of these relics may then, even for ``arbritrarily'' high
primary reheat temperatures of the Universe after an inflationary
epoch, be in accord with current observational
constraints~\cite{BM03,FY02,FY02b}.

  A crucial element in this analysis is the messenger abundance before
decay. This is given, as mentioned earlier, by the thermal freeze-out
of messengers, hence by their annihilation cross-section.  The mass
splitting in the messenger multiplet is generally small, of order
$F_S/M_X \ll M_X$, hence one should in principle consider the various
co-annihilation channels. This task is however left to a future more
refined study; we note that over most of parameter space the inclusion
of co-annihilation channels should modify the relic abundance of the
lightest messenger by at most a factor of order unity since the the
various particles that would co-annihilate have comparable
annihilation cross-sections~\cite{DGP96}.

\subsubsection{Annihilation cross-section}

\paragraph{Annihilation through gauge interactions.}

In the case of $SU(5)$ unification, Dimopoulos {\it et al.}
~\cite{DGP96} have calculated the annihilation cross-section of the
lightest messenger through gauge interactions for
$\mathbf{5}+\mathbf{\overline{5}}$ representations and parametrized it
as:
\begin{equation}
\langle\sigma_{XX^*}v\rangle \,\simeq \, {1\over M_X^2}\left(A+{B\over x}\right),
\label{eq:ann}\end{equation}
with $x\equiv M_X/T$ and $A\simeq -B\simeq 3\cdot10^{-3}$. This
calculation takes into account all annihilation channels (mediated by
gauge interactions) to gauge bosons, Higgses, neutralinos, charginos,
fermions and sfermions, see Ref.~\cite{DGP96}.

\paragraph{Annihilation into two goldstinos.}

  The above calculation neglects annihilation into two goldstinos
$XX^*\rightarrow \tilde G\tilde G$. This latter occurs through a
variety of diagrams: in the $t-$ and $u-$ channels, the annihilation
takes place through the exchange of the fermionic mass eigenstate
partner of $X$. In the $s-$ channel, the annihilation occurs through
the exchange of a graviton, a spurion and other secluded sector scalar
particles. Finally, annihilation also occurs through four-point
contact interactions $XX^*\tilde G\tilde G$. These various
contributions are triggered by various operators in the supergravity
Lagrangian, taking into account the goldstino component of the
gravitino $\Psi_\mu$ and the fermionic spurion $\psi_S$ fields after
supersymmetry breaking as follows:

\begin{eqnarray}
&&\Psi_\mu = i \sqrt{\frac23}\; 
\frac{\partial_\mu \tilde G}{m_{3/2}} + \dots \label{eq:corresp1}\\
&&\psi_S = \frac{F_S}{F} \tilde G +\dots  \label{eq:corresp2}\ .
\end{eqnarray}
For instance, the four-point contact interactions $XX^*\tilde G\tilde
G$ derive from the gravitino mass term in the supergravity Lagrangian
after expanding the exponential of the K\"ahler function to second
order in the scalar around their vev. The Yukawa coupling between the
goldstino and the hidden sector SUSY breaking scalars $Z_i$, which
enters the $s-$channel exchange diagrams, also derives from this
expansion. The coupling $XX^*Z_i$ derives from the scalar potential
trilinear couplings. Finally, the coupling between $X$, its fermions
mass eigenstate partner and $\tilde G$ is obtained directly from the
supergravity Lagrangian coupling between the gravitino and a pair of
fermion-boson partners, taking the appropriate linear combination to
express it in terms of the mass eigenstates after SUSY breaking.

The annihilation cross-section into two goldstinos must be calculated
with care since some leading high energy contributions are expected to
cancel out ~\cite{BR88,G96,G98}. It will be important to distinguish
between the cases where the spurion mass $M_S$ is larger or smaller
than that of the lightest messenger.  Both configurations are
dynamically possible: for instance, one finds in the simplest
models~\cite{GMSB} that $M_X^2 = (\kappa - \sqrt{3} \lambda )(
\kappa/\lambda^2) M_S^2$, where $\lambda$ and $\kappa$ denote
respectively the spurion self-coupling and its coupling to the
messenger fields in the superpotential, and where we have neglected
here the effect of the spurion coupling to the secluded sector fields
following the study of~\cite{DDR97} for the stability and local minima
conditions. In this case the spurion is heavier than the lightest
messenger when $\sqrt{3} \le \kappa/\lambda \lsim 2.2$ and lighter
when $\kappa/\lambda \gsim 2.2$.

In the light scalars and non-relativistic limit, $s\sim
4M_X^2(1+3/x)\gg m_{Z_i}^2$ with $m_{Z_i}$ mass of the secluded sector
scalar $Z_i$, and $x=M_X/T$ as above, the cross-section reads:
\begin{equation}
\langle\sigma_{XX^*\rightarrow \tilde G\tilde G} v\rangle =
\left({F_S\over F}\right)^4 {1 \over 32\pi M_X^2} \left(1 - {15\over 4
x} \right).
\label{eq:ann-gold-hi}
\end{equation} 
In effect, performing an expansion in $F_S/ s$ in the matrix element,
one finds that both terms of order 0 and 1 cancel among the various
contributions, yielding a cross-section $\propto F_S^4$. Its high
energy limit $s\gg M_X^2,m_{Z_i}^2$ is actually that of
spurion-spurion annihilation into two goldstinos~\cite{G98}.  In
practice, Eq.~(\ref{eq:ann-gold-hi}) applies if the spurion is much
lighter than the lightest messenger since its coupling to $XX^*$
dominates that of secluded sector scalars as a result from the
tree-level coupling in the superpotential between $S$, $\Phi_M$ and
$\overline\Phi_M$. However, in this limit one must also account for
annihilation of $X,X^*$ into spurions; this will be discussed further
below.

If secluded sector scalars are heavier than the lightest messenger,
the limit $s\ll m_{Z_i}^2$ applies and in this case, one can neglect
the $s-$channel exchange of these scalars. The cancellation between
the various diagrams occurs only to order 0 in $(F_S/s)$, leaving a
cross-section $\propto F_S^2$:
\begin{equation}
\langle\sigma_{XX^*\rightarrow \tilde G\tilde G}v\rangle\,\simeq\,{1\over
8\pi}{F_S^2M_X^2 \over F^4} \left( 1 - {3\over 2 x}  \right) .
\label{eq:ann-gold-lo}
\end{equation}
The $s-$channel exchange graphs of the secluded sector scalars are
suppressed by $s/m_{Z_i}^2$.  Note that the term $F_S$ in these
expressions should be understood as the mass squared difference
between the fermion and boson components of the lightest messenger
multiplet, rather than as the {\it vev} of the auxiliary component of
$S$. These two quantities differ if the superpotential includes a
coupling constant $\kappa$, $W\supset \kappa S\Phi\overline{\Phi}$;
our choice here is $\kappa=1$.

  The annihilation cross-section in the heavy spurion limit
[Eq.~(\ref{eq:ann-gold-lo})] violates the unitarity bound $\langle
\sigma_{\rm ann}v\rangle \lesssim 8\pi / M_X^2$~\cite{GK90} (in the
non-relativistic regime) for $M_X \gtrsim 1.63\cdot 10^{7}{\rm
GeV}\,(m_{3/2}/{\rm 1\,keV})^{2/3}$. Beyond this limit the effective
Lagrangian is no longer valid, and one expects sizeable contributions
from multi-goldstinos production.  Hence the results obtained
hereafter in the region where the unitarity bound is violated are
highly uncertain and model-dependent. In what follows, we assume that
the cross-section saturates at the unitarity limit in this region $M_X
\gtrsim 1.6\cdot 10^{11}{\rm GeV}\,(m_{3/2}/{\rm 1\,GeV})^{2/3}$ if
the spurion is heavier than the lightest messenger. We also consider
the other possible limit in which the cross-section follows
Eq.~(\ref{eq:ann-gold-hi}), so that the comparison of these two cases
will allow us to assess the impact ofthe above effects on the relic
gravitino abundance.

 Since phenomenology requires that $M_X \sim F_S/10^5\,{\rm GeV}$ (see
Section~II.A), it is easy to see that annihilation into goldstinos
dominates the cross-section in the heavy scalars limit for $M_X
\gtrsim 3\cdot10^6\,{\rm GeV}\,(m_{3/2}/1\,{\rm keV})^{2/3}$. Hence
the inclusion of this channel in the present calculation modifies
rather drastically the relic abundance of the lightest messenger in
this part of parameter space.

\paragraph{Annihilation into two spurions.}

If $S$ is lighter than $X$, there is no problem associated with
unitarity, and one can safely use Eq.~(\ref{eq:ann-gold-hi}) all
throughout parameter space. One must nonetheless account for
$XX^*\rightarrow SS$ annihilation, whose cross-section reads
\begin{equation} \label{sigXXSS}
\langle
\sigma_{XX^*\rightarrow SS} v\rangle = {1 \over 64 \pi M_X^2}
\left( 1 - {1 \over x} + ( -{3\over 2} + {23 \over 4 x} ) r \right)
\end{equation}
to first order in $x^{-1}$ and in $r (\equiv M_S^2/M_X^2)$, and where
we neglected for simplicity the contribution of a $\lambda S^3$ term
in the superpotential, assuming that $\lambda \ll \kappa \simeq 1$.

This cross-section is comparable to the annihilation cross-section
through gauge interactions given in Eq.~(\ref{eq:ann}) and results in
the decrease of the relic abundance of the lightest messenger by a
factor of order 2. For direct GMSB models in which $F_S\sim F$, the
annihilation channel into goldstinos becomes dominant and must be
taken into account.

\subsubsection{Relic abundance}

  Freeze-out of the lightest messenger annihilations occurs at a value
$x_f$:

\begin{equation}
x_f \simeq \log\left[Q_f\left(1 + {B/A\over
\log(Q_f)}\right){1\over\sqrt{\log(Q_f)}}\right],\label{eq:xf}
\end{equation}
where $Q_f\simeq 6.1\cdot10^{10}(M_X/10^6\,{\rm GeV})^{-1}A$, and the
values of $A$ and $B$ accounts for the various possible channels
depicted above. In terms of this freeze-out value $x_f$, the relic
abundance is then given by:

\begin{equation}
Y_X \simeq 2.1\cdot10^{-14}\left({M_X\over 10^6\,{\rm
GeV}}\right){x_f\over A + B/2x_f}.\label{eq:Y_X}
\end{equation}

In the case of messenger sitting in
$\mathbf{10}+\mathbf{\overline{10}}$ representations of $SU(5)$, the
relic abundance is expected to be similar to the above to within a
factor of a few, since the lightest messenger carries hypercharge. For
simplicity, we thus assume that its relic abundance is also given by
Eq.~(\ref{eq:Y_X}) above.

Finally, in the case of $SO(10)$ grand unification, the lightest
messenger is a singlet under the standard model gauge interactions. As
argued in Ref.~\cite{LMJ05}, it can annihilate through one-loop
diagrams (which dominate the exchange of tree level GUT mass bosons
considered in Ref.~\cite{DGP96}) and into two goldstinos at tree level
as above. This case has been discussed in some detail in the low $M_X$
region in Ref.~\cite{LMJ05}.  In Section~\ref{sec:5} we sketch briefly
the parameter space of $SO(10)$ GMSB scenarios using order of
magnitude estimates of these diagrams.  In the main discussion of this
paper, we thus focus on $SU(5)$ grand unification with messengers
either in $\mathbf{5}+\mathbf{\overline{5}}$ or
$\mathbf{10}+\mathbf{\overline{10}}$ representations.

The cosmological scenario we have in mind is the following. As the
Universe reheats to high temperature after inflation, radiation along
with gravitinos and messengers are produced. As the temperature
decreases, the lightest messenger annihilations cease and its
abundance freezes-out. This non-relativistic lightest messenger may
come to dominate the energy density if its decay to the visible sector
is sufficiently delayed and its relic abundance sufficiently large. In
particular, messengers come to dominate the energy density when the
background temperature
\begin{equation}
T_{\rm dom}\simeq {4\over 3}M_XY_X\label{eq:Tdom}
\end{equation}
(provided $T_{\rm dom}\gtrsim T_{\rm dec}$, with $T_{\rm dec}$ the
temperature at which messengers decay, see below). During decay of the
lightest messenger, gravitinos and possibly sparticles may be
produced, but the pre-existing gravitino and NLSP abundances are
diluted by entropy production. Finally, at late cosmological times,
the NLSP decays to the gravitino. In some cases the NLSP may decay
before the lightest messenger. The final gravitino abundance is then
the sum of gravitinos produced by sparticle and messenger interactions
at early times and diluted later by the appropriate factor due to
messenger decay, plus gravitinos produced during the secondary
reheating induced by messenger decay as well as gravitinos produced in
the decay of the NLSP. Then, for a given messenger decay width, which
characterizes the dilution factor, one may find in the $m_{3/2}-M_X$
plane the region where satisfactory abundances for gravitinos are
found.

If $T_{\rm dom} < T_{\rm dec}$, the lightest messenger never comes to
dominate the energy density and no entropy production ensues. This is
what has been generally assumed in previous studies that derived
upper limits on the post-inflationary reheating temperature from 
the upper limit $\Omega_{3/2}<1$, e.g.~\cite{TMY93,CHKL99}.
This case is discussed in Section~IV.A.1.

\section{Gravitino production}\label{sec:3}

\subsection{Production channels}

The fractional contribution to the present critical density of
non-relativistic gravitinos $\tilde G$ with number-to-entropy ratio
$Y_{3/2} = n/s$ is given by \be \Omega_{3/2}\,h^2 = 2.81\cdot 10^8
\left({m_{3/2}\over 1\,{\rm GeV}}\right)\, Y_{3/2} \ee where $h$ is
the Hubble constant in units of 100 km$\,$s$^{-1}\,$Mpc$^{-1}$. The
number-to-entropy ratio $Y_{3/2}$ is found by following the Boltzmann
equation describing gravitino production:
\begin{equation}
{{\rm d}Y\over {\rm d}T}\,=\,{1\over sHT}\left(\sum_i \langle 
\Gamma_{i\rightarrow \tilde G+\ldots}n_i\rangle +
\sum_{i,j}\langle\sigma_{i+j\rightarrow\tilde G+\ldots}v_{ij}n_in_j\rangle\right),
\end{equation}
which includes production by sparticle decays $\Gamma_{i\rightarrow
\tilde G +\ldots}$ and scatterings
$\langle\sigma_{i+j\rightarrow\tilde G+\ldots}v_{ij}n_in_j\rangle$,
neglecting three-body and higher order interactions. In principle, one
should take into account gravitino annihilation as well. However it is
sufficient to approximate possible gravitino losses by imposing that
the gravitino number-to-entropy ratio never exceeds its thermal
equilibrium value $Y_{\rm eq}\simeq 3.7\times
10^{-3}(g_\star/230)^{-1}$, with $g_\star$ the number of relativistic
degrees of freedom in the thermal bath\footnote{Note that this
equilibrium abundance only includes spin 1/2 goldstinos, with the
population of 3/2 components of the gravitino assumed to be
negligible.}.

The production of gravitinos in the early Universe is dominated by the
production of the helicity $\pm1/2$ component if $m_{3/2}\ll M$, where
$M$ denotes the mass scale of particles leading to gravitino
production. This helicity $\pm 1/2$ component is related to the
goldstino through Eq.~(\ref{eq:corresp1}), associated with the
breaking of local SUSY, where $\Psi_\mu$ and $\tilde G$ denote
respectively the helicity $\pm1/2$ gravitino components and the
goldstino~\cite{F79}.  In scenarios of gauge mediated SUSY breaking in
which $m_{3/2}\lesssim 1\,$GeV$\ll M_{\rm SUSY}\sim 1\,$TeV, the
correspondence $\Psi_\mu \sim i \sqrt{2/3}\;\partial_\mu\tilde
G/m_{3/2}$, is generically satisfied.  The production of goldstinos
may then be decomposed into the contributions from particle
scatterings, decay of particles before the freeze-out of the NLSP, and
the possible contribution of NLSP decay.

The decay and scattering contributions of visible sector fields have
been calculated in various studies using the above gravitino-goldstino
equivalence (see e.g., Ref.~\cite{M95} for a detailed discussion and
references). Later on it has been shown that the contribution of
messengers is quite significant~\cite{CHKL99}. Indeed messengers
couple directly to the goldstino $\tilde G$ through the superpotential
term $S\Phi_M\overline\Phi_M$ and the fraction of goldstino $\tilde G$
comprised in the fermionic component of $S$, Eq.~(\ref{eq:corresp2}).
We use the decay widths and cross-sections for messengers interactions
leading to gravitino production given in Ref.~\cite{CHKL99}.

  These authors have also argued that the goldstino decouples from the
visible sector fields at energies $M_X\lesssim E \lesssim \sqrt{F}$
claiming that the effective gravitino-particle-sparticle vertex is
induced by loop diagrams involving messengers~\cite{Lee00}. However
this analysis has been performed in the limit of global supersymmetry
and it thus ignores the tree level fermion-sfermion-goldstino
$-(1/\sqrt{2}m_{\rm
Pl})g_{ij^*}\partial_\mu\Phi^{*j}\overline{\chi}_R\gamma^\nu\gamma^\mu\Psi_\nu
+ {\rm h.c.}$ and gauge boson-gaugino-goldstino $(i/4m_{\rm
Pl})\overline\Psi_\mu\sigma^{\rho\sigma}
\gamma^\mu\lambda_L^{(a)}[F_{\rho\sigma}^{(a)}+\tilde
F_{\rho\sigma}^{(a)}]$~\cite{WBook} interaction terms which appear in
local supersymmetry. Therefore the goldstino does not decouple from
visible sector fields at energies $M_X\lesssim E \lesssim \sqrt{F}$,
at variance with Ref.~\cite{CHKL99}. Nevertheless, it is true,
following Refs.~\cite{Lee00,CHKL99} that production channels should
include the loop-induced messenger contribution to the effective
particle-sparticle-goldstino vertex at energies $E\lesssim M_X$. This
is done in the present study; we do not include thermal corrections to
the cross-sections since they are found to be
negligible~\cite{ENOR96}.

The various contributions to $Y_{3/2}$ have different temperature
dependences. The largest fraction of the decay contributions results
from cosmic epochs when the temperature falls below the mass of the
corresponding particles. In contrast the dimension-5 operators
associated to visible sector particle -- sparticle scatterings are
most effective at high temperatures; the contribution from messengers
peaks at temperatures $T\sim M_X$. A simple fit to the results of
Ref.~\cite{CHKL99}, with the modifications according to the remarks above,
gives an estimate of the amount of gravitinos
produced by scatterings and decays of sparticles and messengers for a
reheating temperature $T_{RH}$:
\begin{eqnarray}
\Omega_{3/2}h_{70}^2 & \sim & 2\cdot 10^{-4}\left({m_{3/2}\over 1\,{\rm
GeV}}\right)^{-1}\Biggl[ 0.6\left({M_{3}\over 10^3\,{\rm GeV}}\right)^3
\nonumber\\ & & + 2.4 \left({M_{3}\over 10^3\,{\rm GeV}}\right)^2\left({T_{\rm
RH}\over 10^5\,{\rm GeV}}\right)\Biggr]\nonumber\\ & & +\,
2\left({m_{3/2}\over 1\,{\rm GeV}}\right)^{-1}\Biggl[
3\left({M_{3}\over 10^3\,{\rm GeV}}\right)^4\left({M_X\over 10^5\,{\rm
GeV}}\right)^{-1}  \nonumber\\ & & + 0.2\left({M_{3}\over 10^3\,{\rm
GeV}}\right)^2\left({M_X\over 10^5\,{\rm
GeV}}\right)\Biggr]\,.
\label{eq:CHKL}
\end{eqnarray}
The first two lines correspond to sparticle decays and
particle-sparticle scatterings, respectively,
while the last two correspond to interactions
involving messenger fields. Hence the latter should only be included
if $T_{\rm RH}\gtrsim M_X$.  In these equations, $M_3$ denotes the
gluino mass scale, which controls the amount of gravitinos
produced. Indeed, the coupling between gauginos, gauge bosons and
golstinos dominates the particle-sparticle-goldstino contributions at
high temperatures and scales as the gauge coupling constant; its
contribution is represented by the second term on the r.h.s. of
Eq.~(\ref{eq:CHKL}). Decay contributions to $\Omega_{3/2}h^2$ scale as
$M^3$, with $M$ the mass of the decaying sparticle, see the first term
on the r.h.s. of Eq.~(\ref{eq:CHKL}). The gluino appears in this term
due to the larger degree of freedom
for colored particles than for color singlets as well as due to
colored sparticles being heavier than color singlets in
the ratio $\alpha_3/\alpha_2$ or $\alpha_3/\alpha_1$ [see
Eq.~(\ref{eq:GMSBmass})]. 
Strictly speaking this relation applies at
the messenger scale and must be corrected by renormalization group
running at lower energy scales. Nevertheless the mass spectrum of GMSB
models is dominated by squarks and gluinos whose masses are comparable
at scale $M_X$. As regards the contribution of messengers, given in
Eq.~(\ref{eq:CHKL}) for $T_{\rm RH}\gtrsim M_X$, the gluino mass
appears only in the combination $(4\pi/\alpha_3)M_3 \sim F_S/M_X$,
i.e. as the normalization of $F_S$ as a function of $M_X$ and the
electroweak scale.

 We provide Eq.~(\ref{eq:CHKL}) in order to assist the reader in
interpreting the figures that follow. The results shown below are
obtained from the integration of the Boltzmann equation. Also recall
that Eq.~(\ref{eq:CHKL}) does not take into account all contributions
to the gravitino abundance. One must notably add the contribution of
gravitinos produced by annihilations of the lightest messenger, as
well as lightest messenger decay and NLSP decay.

The NLSP decays into final states including one gravitino with width:
\begin{equation}
\Gamma_{\rm NLSP\rightarrow \tilde G} \simeq
\frac{1}{48\pi}\frac{M_{\rm NLSP}^5}{m_{3/2}^2m_{\rm Pl}^2} \ .
\label{eq:G_NLSP}
\end{equation}
The background temperature at NLSP decay is then $T_{\rm
NLSP\rightarrow \tilde G}\simeq 5\, {\rm MeV}\,(M_{\rm NLSP}/100{\rm
GeV})^{5/2}(m_{3/2}/1{\rm MeV})^{-1}$. This decay occurs late and
consequently big-bang nucleosynthesis constraints on hadronic or
electromagnetic energy injection at time $\sim 10^{-2}-10^8\,$sec lead
to the exclusion of a significant part of parameter space, as will be
seen in the following. Since one NLSP produces one gravitino, the
gravitino yield of NLSP decays in terms of entropy density is simply
$Y_{\rm NLSP}$.

  In principle, the NLSPs result from a freeze-out of thermal
equilibrium, hence $Y_{\rm NLSP}$ can be calculated in the same way as
the relic abundance of the lightest messenger. However, one must not
forget the possible production of NLSPs during the decay of the
lightest messenger. If the decay temperature of the latter is larger
than the NLSP freeze-out temperature, this contribution is washed out
by NLSP annihilations.  However if decay occurs later, NLSPs are
regenerated to a level which depends on the time of messenger decay
and on the annihilation cross-section of the lightest messenger. In
more detail, if the lightest messenger decays to NLSPs before NLSPs
decay in turn to gravitinos, the NLSPs have time to annihilate. In
Ref.~\cite{FY02}, a procedure was outlined to calculate the number
density of NLSPs remaining after these further annihilations. If the
lightest messenger decays to NLSPs after pre-existing NLSPs have
decayed to gravitinos, the calculation of the final number of
gravitinos produced is more involved and necessitates the integration
of coupled Boltzmann equations. For simplicity, we assume that all
NLSPs produced in messenger decay then decay instantaneously to
gravitinos without annihilating. This maximizes the number of
gravitinos produced hence reinforces the constraints from relic
density arguments, and, in this sense, this assumption is
conservative. Keeping track of NLSP annihilation before decay to
gravitinos would not affect our results significantly as it is
marginal in most of parameter space~\cite{FY02}.

Finally there exist other potential channels of gravitino
production. One is that of helicity $\pm3/2$ production, which for
large gravitino mass and high reheating temperature may become
important. The helicity $\pm3/2$ modes interact with gravitational
strength only and are produced by interactions in the thermal bath in
abundance $\Omega_{3/2}^{\pm3/2}h^2 \sim (m_{3/2}/1{\rm GeV})(T_{\rm
RH}/10^{14}{\rm GeV})$~\cite{M95}. Therefore this contribution does
not dominate in most of the $m_{3/2}-M_X-T_{\rm RH}$ parameter
space. We do not take into account possible non-thermal production
channels of helicity $\pm 3/2$ gravitinos during
inflation~\cite{non-th}. The amount of gravitinos produced in this way
depends strongly on the underlying model of inflation, eventhough it
may exceed the amount of helicity $\pm3/2$ modes produced by
scatterings in the thermal bath in particular models.

\subsection{Gravitino Dilution}

The delayed decay of non-relativistic messengers may have dramatic
consequences on the abundance of any pre-existing species, such as
gravitinos.  In case delayed decay results in the temporary
matter-domination by messenger rest mass, i.e. $\rho_X\gg\rho_{\rm r}$
where $\rho$ denote energy densities and subscripts $X$ and ``${\rm
r}$'' refer to messenger and radiation, respectively, entropy
production is significant and results in the severe dilution of any
pre-existing number-to-entropy ratio $Y$. In this case the
post-messenger-decay cosmic radiation temperature $T_{\rm dec}^>$ is
substantially larger than the pre-decay temperature $T_{\rm dec}^<$,
akin of a second reheat.  Approximating decay to be instantaneous when
the Hubble scale equals the decay width of the lightest messenger,
$H\approx \Gamma_X$, one finds

\begin{equation}
T_{\rm dec}^> \approx (g_>\pi^2/90)^{-1/4}\sqrt{\Gamma_Xm_{\rm Pl}},
\label{eq:Tda}
\end{equation}
where $g_>$ denotes the number of relativistic d.o.f. at temperature
$T_{\rm dec}^>$. If the particle decays into the visible and into an
invisible sector, the decay width in Eq.~(\ref{eq:Tda}) above should
be multiplied by $B_{\rm visible}$, with $B_{\rm visible}$ the
branching ratio into visible sector particles. This may be of
relevance notably when the lightest messenger decays into visible
sector particles and into gravitinos which do not share their energy
density with the visible sector afterwards.

  By equating the pre- and post-decay energy densities, the pre-decay
radiation temperature $T_{\rm dec}^<$ is obtained in terms of $T_{\rm
dec}^>$ and $T_{\rm dom}$ [see Eq.(\ref{eq:Tdom})] at which $X$ comes
to dominate the energy density, as:

\be T_{\rm dec}^< \approx T_{\rm dec}^>\,\, {\rm min}\,\left[1\,
,\,\left(\frac{g_>}{g_<}\right)^{1/3}\left({T_{\rm dec}^>\over T_{\rm
dom}}\right)^{1/3}\right].
\label{eq:TbTa}
\ee 

Obviously, if $X$ does not dominate the energy density before
decaying, $T_{\rm dec}^< \approx T_{\rm dec}^>$, while if $T_{\rm
dec}^> \ll T_{\rm dom}$, one finds $T_{\rm dec}^< \ll T_{\rm dec}^>$
and entropy production is very substantial. In effect the ratio of
pre-decay and post-decay entropy densities, gives the entropy release
$\Delta_X\equiv s_>/s_< = g_>T_>^3/g_<T_<^3$:

\be \Delta_X \approx\,{\rm max}\,\left[1\,,\, {T_{\rm dom}\over
T_{\rm dec}^>}\right],
\label{Delta1}
\ee The values of $T_{\rm dec}^>$ and $T_{\rm dom}$ are given in terms
of $Y_X$, $M_X$ and $\Gamma_X$ through
Eqs.~(\ref{eq:Tda}) and (\ref{eq:Tdom}).

Such entropy release dilutes pre-existing densities according to: $Y_>
= Y_</\Delta_X$. Nevertheless, it should be borne in mind that, in
case of high second reheat temperatures $T_{\rm dec}^>$, the
regeneration of diluted species may occur. This effect is taken into
account in our calculations by treating messenger decay as a second
reheat.  Note that substantial entropy release after BBN is
unacceptable, and $T_{\rm dec}^> \gtrsim 1\,$MeV is
required. Eq.~(\ref{eq:Tda}) may thus be employed to infer a fairly
strict lower limit of $\Gamma_X\simge 4.3\times 10^{-25}\,$GeV on
abundant and slowly decaying particle species in the early Universe.

Particularly interesting to cosmology is the case of significant
entropy dilution.  In this limit one may use Eqs.~(\ref{Delta1})
and~(\ref{eq:TbTa}) to derive the entropy dilution factor for an
abritrary species with mass $M_X$, decay width $\Gamma_X$ and
abundance $Y_X$, in the limit $\Delta_X\gg 1$:
\bea \Delta_X & \approx & 0.77\, g_>^{1/4}\, Y_X\, \Gamma_X^{-1/2}\, m_{\rm
pl}^{-1/2}\, M_X\, \nonumber\\ &  \approx & 28\left({M_X\over 10^8{\rm
GeV}}\right)\left({Y_X\over 10^{-10}}\right)\left({\Gamma_X\over
10^{-25}{\rm GeV}}\right)^{-1/2}\left({g_>\over 10}\right)^{1/4},
\label{Delta2} \eea where it is understood that if
$\Delta_X\simle 1$ is found, it ought to be substituted by $\Delta_X =
1$.

If there exists a whole tower of $N$ unstable, but long-lived
particles, with abundances $Y_i$, masses $M_i$ and decay widths
$\Gamma_i$ for particle $i$, the final dilution factor is determined
solely by the properties of the slowest decaying messenger. In
particular, all the equations above may be employed as if any prior
decays had not occurred.

\subsection{Cosmological constraints}

\subsubsection{Hot, warm or cold dark matter}

Collisionless damping during the radiation era leads to the erasure of
power in density fluctuations below a length scale (the free-streaming
scale) that is mostly determined by the time at which dark matter
particles become non-relativistic, or equivalently by their velocity
extrapolated to zero redshift. A particle of mass $m$ that thermally decouples
from the plasma when relativistic has a present-day velocity:

\begin{equation}
\langle v_0 \rangle \approx 0.018\,{\rm km}\,{\rm
s}^{-1}\,(g_{\star,{\rm dec}}/230)^{-1/3}(m/1{\rm keV})^{-1},
\label{eq:v-th}
\end{equation}
with $g_{\star,{\rm dec}}$ the number of d.o.f. at
decoupling. Cosmological data on the power spectrum of density
fluctuations allow to place constraints on the mass of the
particle. One finds $m\gtrsim 1\,$keV from the requirement that the
Universe has reionized by $z\sim 6$~\cite{BHO01} or from the
measurement of the power spectrum in the Lyman~$\alpha$
forest~\cite{DDT03}. If reionization has occurred as early as $z\sim
17$, as suggested by the recent WMAP data~\cite{WMAP}, then a mass
larger than $\sim 10\,$keV seems required~\cite{YSHS03}.

These constraints are important to our analysis and we keep track of
the average velocity extrapolated to $z=0$. Obviously the limits
between hot, warm and cold matter are fuzzy, and we choose to qualify
as warm dark matter particles with velocity $0.0018\,{\rm km}\,{\rm
s}^{-1}\,\leq v_0 \leq 0.054 \,{\rm km}\,{\rm s}^{-1}\,$, corresponding
to particle masses $0.3\,{\rm keV}\,\leq m \leq 10\,{\rm keV}\,$ (for
freeze-out from thermal equilibrium as above). For velocities above
the upper limit or below the lower limit, we mean hot or cold dark
matter respectively. It is important to note that entropy production
after decoupling of the gravitinos cools down these dark matter
particles according to: $\langle v_0 \rangle \rightarrow \langle v_0
\rangle /\Delta_X^{1/3}$.

For gravitinos produced by out-of-equilibrium processes, notably by
the late decay of a massive particle, the above relation between mass
and velocity is modifed. Assuming the outgoing gravitino carries a
momentum of half the mass $M$ of the decaying particle, the present
velocity reads:
\begin{equation} \langle v_0\rangle \approx (M/2m_{3/2})
(3.91/g_{\star,{\rm dec}})^{1/3}(T_0/T_{\rm
dec}),\label{eq:v-dec}\end{equation} with $g_{\star,{\rm dec}}$ the
number of d.o.f. at decay, $T_0$ the present cosmic background
temperature and $T_{\rm dec}$ the temperature at decay. Note that
$T_{\rm dec}$ generally depends on $M$ so that the dependence between
the nature of gravitino dark matter (cold/warm/hot) and the mass of
the decaying particle is not necessarily trivial. For instance, one
can show that a decaying NLSP produces hot/warm dark matter if its
mass $\lesssim 500\,$GeV. 

  If the decay occurs at temperatures sufficiently high that the
gravitino can interact and thermalize, one should rather use
Eq.~(\ref{eq:v-th}). However at temperatures $T\ll100\,$GeV the
gravitino has decoupled from the thermal plasma~\cite{G96}, mainly
because the sparticles have decoupled themselves. Therefore the decay
of the NLSP or of the lightest messenger (if sufficiently late)
generally produces highly relativistic gravitinos.

\subsubsection{Big-bang nucleosynthesis constraints}

  Due to the different channels of gravitino production, one generally
finds that gravitinos are made of two generic sub-populations: one
that has been produced by equilibrium processes and another made of
hot gravitinos produced by out-of-equilibrium decays. It may be that
the latter are so highly relativistic that they form a hot dark matter
component. However their impact on the formation of large scale
structure may be negligible if their contribution to the gravitino
energy density is negligible. In this particular case, constraints
from big-bang nucleosynthesis on extra degrees of freedom may apply
and constrain this population. We take the BBN constraints on extra
degrees of freedom to be $\delta g \leq 1.8$ (corresponding to 1
extra neutrino family allowed)~\cite{BBN-dof}. Gravitinos that are
relativistic at the time of BBN and carry energy density $\rho^{\rm
r}_{3/2}$ contribute to the level $\delta g_{\rm BBN}/g_{\rm BBN}^s
\simeq \rho^{\rm r}_{3/2}/\rho_{\rm r}^s$, with 
$\rho_{\rm r}^s$ the standard radiation energy density at the onset of BBN
with $g_{\rm BBN}^s=10.75$.
Gravitinos that were once in thermal equilibrium or that were produced
by equilibrium processes at temperature $T$ (d.o.f. $g$) carry
characteristic momentum $p_{3/2}\sim 3T_{\rm BBN}(g_{\rm
BBN}^s/g^s)^{1/3}$ at BBN, with $T_{\rm BBN}\sim 1\,$MeV. Assuming
$g^s\sim 230$, these gravitinos are relativistic if $m_{3/2}\lesssim
1\,$MeV and their contribution to the energy density is $\delta
g_{\rm BBN}/g_{\rm BBN}^s \simeq Y^{\rm r}_{3/2}$ with $Y^{\rm
r}_{3/2}=n^{\rm r}_{3/2}/s$, hence it is negligible due to the upper
bound on $Y_{3/2}$ resulting from thermal equilibrium. 

However most relativistic gravitinos at the time of BBN result from
out-of-equilibrium decays, e.g. from NLSP or from the lightest
messenger decay. Given that, immediately after decay the outgoing
gravitinos carry a fraction $B_{3/2}$ of the rest mass energy of the
decaying particle of mass $M$ as kinetic energy, their contribution to
the energy density at the onset of BBN reads: $\delta g_{\rm
BBN}/g_{\rm BBN}^s \simeq B_{3/2} M Y_M (4/3) (g_{\rm
BBN}^s/g^s)^{1/3}/T_{\rm d}$, where $T_{\rm d}$ is the decay
temperature and $Y_M$ the number-to-entropy ratio of the parent at
decay. Here it has been assumed that $T_{\rm d}\simge 4 M Y_M/3$. In
the opposite limit, i.e.  in the case of significant entropy
production at decay, the above relation becomes $\delta g_{\rm
BBN}/g_{\rm BBN}^s \simeq B_{3/2}(g_{\rm BBN}^s/g^s)^{1/3}$ (assuming
$B_{3/2}\ll1$), and it will be this limit which results in the
strongest BBN constraints.

The time of decay of NLSPs to gravitinos is also strongly constrained
by big-bang nucleosynthesis limits on hadronic and electromagnetic
energy injection at times $\sim 10^{-2}\rightarrow 10^8\,$sec. These
constraints have been examined in Refs.~\cite{BBN1,BBN2,BBN4,FST04},
while Ref.~\cite{GGR99} has translated these bounds on the messenger
scale $M_X$ of GMSB models assuming $k=1$ (which is equivalent to
setting an upper bound on $m_{3/2}$). In the present analysis, we use
the latest constraints from hadronic and electromagnetic energy
injection from Ref.~\cite{BBN2}.  

In GMSB scenarios, the NLSP is generically a neutralino (mainly bino)
or a stau. The former decays predominantly into a photon and a
goldstino; the fraction of energy spent with branching ratio $B_{\rm
em}\simeq 1$ and by three body decays into a pair of quarks and
goldstino with hadronic branching ratio $B_{\rm had}\sim 10^{-3}$; if
its decay to $Z$ bosons is not suppressed by phase space,
i.e. $(M_{\rm NLSP}-M_Z)/M_Z\gtrsim 1$, the hadronic branching ratio
$B_{\rm had}\sim 0.15$. For simplicity, we use this latter value,
which is conservative with respect to the constraints
inferred. Concerning the annihilation cross-section of the bino NLSP,
we use the value $\langle\sigma_{\rm NLSP} v\rangle = 10^{-9}\,{\rm
GeV}^{-2}\,(M_{\rm NLSP}/100\,{\rm GeV})^{-2}$, which corresponds to
the bulk region of minimal supergravity~\cite{FST04}. We will comment
on the dependence of our results on the choices made when discussing
the results shown in Fig.~\ref{fig:fast-T} below.

A stau NLSP may produce in its decay electromagnetic and hadronic
showers. About 100\% of the energy is converted to electromagnetically
interacting particles. In 70\% of all decays, a stau NLSP produces
hadrons, but these are mesons whose lifetimes are so short that they
do not have time to interact before decaying if they were emitted at
times $\gtrsim 10^2\,$sec. Hence, we use $B_{\rm had}=0.7$ for stau
decay timescales shorter than $10^2\,$sec and $B_{\rm had}=10^{-3}$
for longer decay times. The stau annihilation cross-section is not as
model dependent as that of the bino, $\langle \sigma_{\rm
NLSP}v\rangle\simeq 10^{-7}\,{\rm GeV}^{-2}\,(M_{\rm NLSP}/100\,{\rm
GeV})^{-2}$. Given its large annihilation cross-section, the stau has
a small relic abundance, and consequently the BBN constraints are
comparatively weaker.

Overall big-bang nucleosynthesis constraints apply to a combination of
$Y_{\rm NLSP}$ (relic abundance) and decay timescale $\propto M_{\rm
NLSP}^{-5}m_{3/2}^2$. Note that for a decay timescale $\tau\sim
10^3\,$sec, interesting modifications to BBN may result~\cite{BBN4}.
We also note that in a very limited part of parameter space of GMSB
theories, the NLSP can be a sneutrino~\cite{GR99}, for which the BBN
constraints would be largely reduced~\cite{KM95,FST04}.

  Finally, since the mass of the NLSP enters the BBN constraints while
the gravitino yield is controled by the gluino mass, it is necessary
to schematize the mass spectrum of GMSB scenarios. We do so by
assuming a mass ratio $M_3/M_{\rm NLSP}\sim 6$~\cite{GR99,DTW97} and
fiducial values $M_{\rm NLSP}=150\,$GeV and $M_3=1\,$TeV. Where
relevant we mention the possible influence of these values on our
results.

\section{Messenger couplings to matter and gravitino dark matter}\label{sec:4}

In this section, we investigate the possible solutions for gravitino
dark matter for various messenger number violating interactions added
to the Lagrangian. As argued in Section~\ref{sec:2}, such interactions
are mandatory if no substantial entropy production occurs at
temperatures $\lesssim M_X$ (other than due to lightest messenger
domination and decay) in order to avoid the cosmological problems that
would result from the stability of the lightest messenger. For
definiteness, we adopt the notations of Ref.~\cite{WBook} including
four component spinors, in the general supergravity Lagrangian.

\subsection{Renormalizable couplings}

\subsubsection{Superpotential couplings}

Renormalizable couplings, beyond those of the required messenger gauge
interactions, may exist, though they are constrained by considerations
of flavor changing neutral currents as well as the potential
development of charge- and color- breaking minima, among other issues.  Dine
{\it et al.}~\cite{DNS97} have analysed viable extensions of the
minimal GMSB scenario in this direction, introducing couplings of the
form \be W \supset y_l^i H_D\Phi^l_M\overline{e}_i + y_q^i
H_DQ_i\overline{\Phi}^q_M\label{WDine},\ee where $\Phi^l_M$ and
$\overline{\Phi}^q_M$ denote lepton- and quark- like messengers.  Here
$\Phi_l$ denotes an SU(2) messenger doublet, the $y_i$'s are Yukawa
couplings with family index $i$, and $H_D$, $\overline{e}$, and $Q_i$
are standard model down-type Higgs, right-handed lepton, and quark
doublet, respectively.

Additional SUSY-breaking mass splittings are generated by these types
of interactions via one-loop contributions yielding, for example,
negative relative mass contributions of order $\delta
m_{\tilde{e}}/m_{\tilde{e}}\approx -10^3y_l^2 F_S^2/M_X^4$ to slepton
masses. Flavor changing neutral currents may place potentially
restrictive limits on such couplings, as due to the experimentally
verified weakness of such processes, mass splittings between 1st- and
2nd- generation sleptons (and squarks) are constrained to be
smaller than $(m^2_{\tilde{e}_1}-m^2_{\tilde{e}_2})/m^2_{\tilde{e}} \simle 10^{-3}$. 
Assuming conservatively, $y^1_l\neq 0$ and $y^2_l=y^3_l=0$ one
may thus infer a limit $y^1_l\simle (M_X/10^8\GeV)$ on this extra
Yukawa coupling.

Interactions induced by the superpotential Eq.~(\ref{WDine}) also
induce the decay of messengers, in particular $X\rightarrow
\tilde{H}^- e^+$ assuming $X$ carries the same gauge charges as a
$\tilde\nu_L$ (see Section~\ref{sec:2}). Hence the decay width
$\Gamma_X = y^2M_X/8\pi$. Though the limit on Yukawa couplings as
inferred above may be quite severe, entropy production due to delayed
messenger decay is absent when terms of Eq.~(\ref{WDine}) are included
into $W$, unless the extra Yukawa coupling is extremely small,
$y\lesssim 10^{-15} (M_X/10^7\,{\rm GeV})^{3/2}$, assuming $SU(5)$
grand unification.

\begin{figure*}
  \centering
  \includegraphics[width=0.9\textwidth,clip=true]{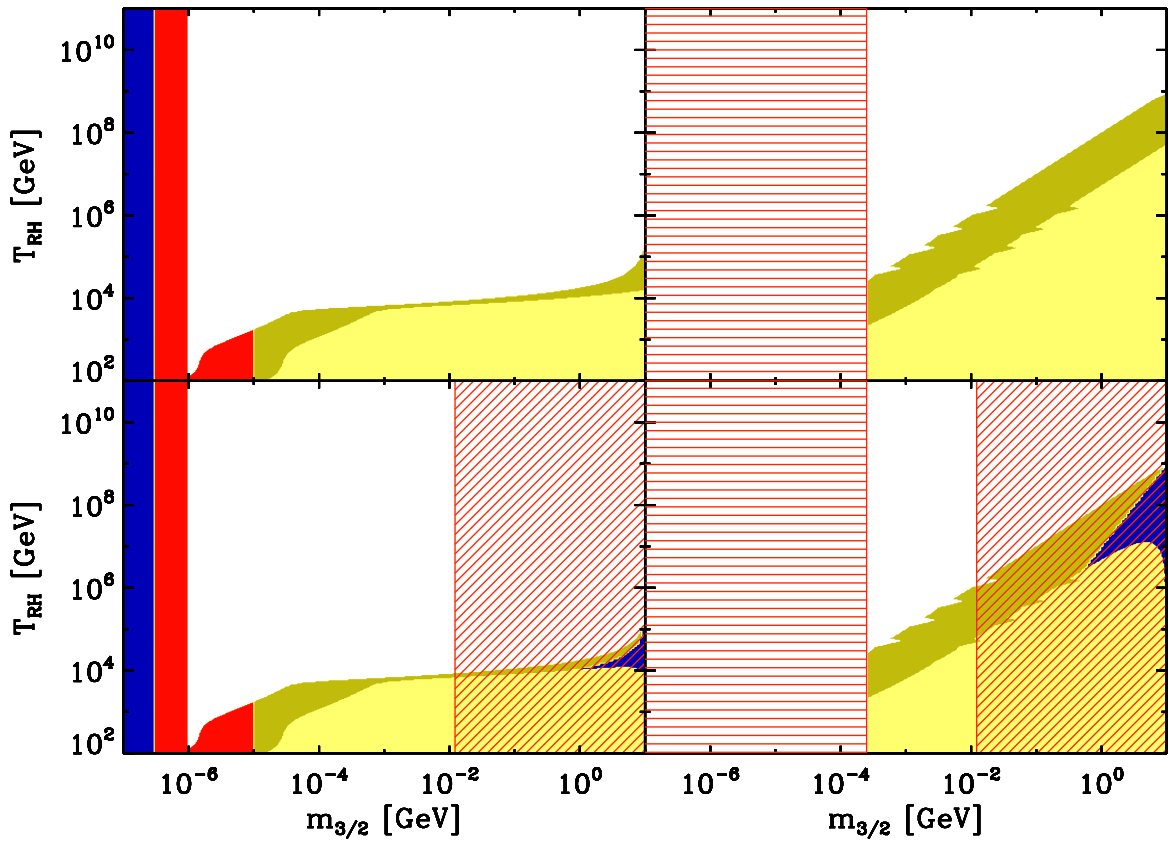}
  \caption[...]{Contours of $\Omega_{3/2}$ in the plane $T_{\rm
  RH}-m_{3/2}$ for ``fast'' messenger decay, as discussed in the
  text. White corresponds to $\Omega_{3/2}>1$; increasingly lighter
  shaded areas (blue, red and green respectively) from left to right
  (in the left panel) correspond to hot, warm or cold dark matter
  respectively with $0.01\leq\Omega_{3/2}\leq1$. The lightest zone
  (yellow) indicates $\Omega_{3/2}<0.01$. The area shaded (in red) by
  lines oriented NE-SW is the zone excluded by BBN constraints on NLSP
  decay. The zone shaded with horizontal lines is theoretically
  excluded as it corresponds to $F_S>F$. The NLSP is bino-like in the
  lower panels, and stau-like in the upper panels. The messenger mass
  scale is $M_X=10^5\,$GeV in the left panels and $M_X=10^{10}\,$GeV
  in the right panels. }
\label{fig:fast-T}
\end{figure*}

  It is nevertheless instructive to study the influence of such
``fast'' decay of the lightest messenger on the possiblity of having
gravitino dark matter. As mentioned earlier, this case (without
entropy production) has been implicitly assumed in previous studies
that have drawn upper bounds on the post-inflationary reheating
temperature from the upper bound on the gravitino density
$\Omega_{3/2}<1$~\cite{TMY93,CHKL99} (and references therein).  In
order to provide a point of comparison with this previous litterature,
we plot in Fig.~\ref{fig:fast-T} the results of the calculation of
$\Omega_{3/2}$ in the plane $T_{\rm RH}-m_{3/2}$, using the techniques
developed in the previous section.  We assume ``fast'' decay with
width $\Gamma_X \sim 10^{-9}M_X$ corresponding to $y\sim 10^{-4}$,
which ensures that phenomenological constraints are satisfied for all
values of $M_X$.  The results shown are insensitive to the exact value
of $y$, provided it is not so tiny that substantial entropy production
would occur.

The shade (color) coding in this figure and all subsequent figures is
as follows: lightest (yellow) corresponds to $\Omega_{3/2}<0.01$ (no
gravitino problem but no dark matter), and the increasingly darker
(respectively green, red and blue) areas indicate respectively the
regions of cold, warm and hot dark matter in which
$0.01<\Omega_{3/2}<1$. The area shaded by lines oriented NE-SW at the
right of each figure corresponds to the region excluded by BBN
constraints on NLSP to gravitino decay. White color indicates
$\Omega_{3/2}>1$, i.e., overclosure of the Universe by
gravitinos. Finally the area marked with horizontal lines is
unphysical as it corresponds to $F_S>F$ ($k>1$).

  We choose $0.01$ and $1$ as lower and upper bounds respectively to
delimit where the gravitino can account for dark matter, eventhough
cosmological data restrict this to a much smaller range. However the
calculations presented here contain intrinsic uncertainties of factors
of a few that were mentioned in the previous sections, and therefore
the green, red and blue areas should be understood as indicative of
the region in which one can find solutions for gravitino dark matter.

  As indicated in the caption of Fig.~\ref{fig:fast-T}, the left
panels correspond to $M_X=10^5\,$GeV while the right panels correspond
to $M_X=10^{10}\,$GeV (for which the condition $k\leq1$ translates in
$m_{3/2}\gtrsim 250\,$keV). The upper and lower panels correspond
respectively to a stau and a bino NLSP. As anticipated in the previous
Section, the BBN constraints on hadronic and electromagnetic energy
injection do not apply to the stau at these small values of $m_{3/2}$,
as a result of the low stau relic abundance. In fact, for a stau of
mass $M_{\rm NLSP}\simeq 150\,$GeV, a gravitino as heavy as
$m_{3/2}\sim 10\,$GeV is allowed by BBN~\cite{FST04}. However, the
constraints are quite stringent for the case of the bino NLSP, and
result in an upper bound $m_{3/2}\lesssim 10\,$MeV for $M_{\rm
NLSP}=150\,$GeV. At a fixed value of the NLSP relic abundance, the BBN
constraints give an upper bound on the decay timescale; hence, the
above limit on $m_{3/2}$ scales as $M_{\rm NLSP}^{5/2}$, see
Eq.~(\ref{eq:G_NLSP}). The BBN limit on $m_{3/2}$ evolves as follows
with respect to the bino annihilation cross-section; for reference, we
recall that the cross-section used in the calculations reported in
Fig.~\ref{fig:fast-T} is $\langle\sigma_{\rm NLSP}v\rangle\simeq
\sigma_0\,(M_{\rm NLSP}/100\,{\rm GeV})^{-2}$ with
$\sigma_0=10^{-9}\,$GeV$^{-2}$ and $M_{\rm NLSP}=150\,$GeV. If
$\sigma_0$ is decreased by a factor 10 to 100, the upper bound on
$m_{3/2}$ shifts to $\sim 100\,$MeV; if, conversely, the cross-section
is increased by a factor 10 to 100, the upper bound on $m_{3/2}$
shifts to $\sim3\,$MeV. As mentioned in the previous section, we have
implicitly assumed a branching ratio to hadronic decay $B_{\rm
had}=0.15$; if the bino is nearly degenerate in mass with $Z$, the
hadronic decay mode is suppressed, with a value possibly as small as
$B_{\rm had}\sim 10^{-3}$. In this case, for $\sigma_0$ chosen as
above, the bound on $m_{3/2}$ would be $\sim 100\,$MeV, increasing to
$\sim 10\,$GeV if $\sigma_0$ is increased by a factor 100, and
remaining constant if $\sigma_0$ is decreased by a factor as large as
100.  Overall the BBN constraints result in a bound $m_{3/2}\lesssim
10\,{\rm MeV}\rightarrow 1\,{\rm GeV}$ depending on the bino mass,
annihilation cross-section and hadronic branching ratio.

 Figure~\ref{fig:fast-T} illustrates the so-called light gravitino
problem: if $m_{3/2}\simge 1\,$keV, gravitinos overclose the Universe
and/or disrupt BBN unless the reheating temperature is low, $T_{\rm
RH}\sim {\rm min}\left[10^8 m_{3/2}(M_3/1{\rm
TeV})^{-2},\,M_X/10\right]$, and gravitinos are not too heavy,
$m_{3/2}\simle 10\,{\rm MeV} - 1\,$GeV.  Although it is not impossible
to achieve such small reheating temperatures, either by low-scale
inflation or a late phase of thermal inflation, it is not particularly
attractive as it puts further non-trivial requirements on the model
and may pose problems for a successful genesis of baryon or lepton
asymmetry. Moreover it is necessary to tune the reheating temperature
to the gravitino and/or messenger mass, eventhough these quantities
derive from sectors of the theory that are {\it a priori} unrelated.

  The region $m_{3/2}\lesssim 1\,$keV is devoid of constraints on
$T_{\rm RH}$ since the gravitino is so light that even at thermal
equilibrium, it cannot overclose the Universe. However such light
gravitinos make up dark matter that is too warm to reproduce existing
data on the large scale structures. Heavier gravitinos $m_{3/2}\gtrsim
10\,$MeV are excluded by big-bang nucleosynthesis constraints if the
NLSP is a bino. Since NLSPs decay at time $\tau_{\rm NLSP}\sim 6\times
10^4\,{\rm sec}\,(M_{\rm NLSP}/100{\rm GeV})^{-5}(m_{3/2}/1{\rm
GeV})^2$, the heavier the gravitino the later and the more constrained
the decay.  Figure~\ref{fig:fast-T} differs from those shown in
Refs.~\cite{TMY93,CHKL99} because we have included constraints from
GMSB phenomenology (namely $k\leq 1$) as well as updated constraints
from BBN and structure formation, and a more accurate calculation of
the gravitino relic abundance.  Overall, one finds that the range of
allowed $m_{3/2}$ is severely restricted when compared to previous
studies. 

In particular, the present conclusion is at variance with
Ref.~\cite{CHKL99} which argued that for a sufficiently small
messenger mass scale $M_X\sim 10^5\,$GeV and sufficiently large
gravitino mass $m_{3/2}\gtrsim 2\,$GeV, it is possible to find
solutions to the gravitino problem for arbitrarily high reheating
temperatures. The discrepancy with \cite{CHKL99} is tied to the
neglect in that study of the SUGRA induced MSSM particles contribution
to the gravitino abundance at large reheating temperatures, as
discussed in Section~III.A., as well as of the big-bang
nucleosynthesis constraints on NLSP decay.  As can be seen in the left
panels of Fig.~\ref{fig:fast-T}, the upper bound on $T_{\rm RH}$ does
indeed shift upwards, albeit for larger $m_{3/2}$ than expected in
~\cite{CHKL99}, i.e. $m_{3/2}\gtrsim 10\,$GeV for $M_X\sim
10^5\,$GeV. This $m_{3/2}$ region is however forbidden by BBN
constraints on energy injection, for both stau and bino NLSP. If the
gluino mass scale is smaller than the fiducial value of $1\,$TeV, say
$M_3=300\,$GeV, this region of parameter space where the bound on
$T_{\rm RH}$ is relaxed, moves to smaller $m_{3/2}$,
i.e. $m_{3/2}\gtrsim 0.4-1\,$GeV. However, the BBN constraints also
move to smaller $m_{3/2}$ since the NLSP mass is reduced as the gluino
mass. Finally, since the MSSM particle contribution to $\Omega_{3/2}$
contains a dependence on $m_{3/2}$ and $T_{\rm RH}$, we find that
$T_{\rm RH}$ is always bounded from above due to the BBN bound on
$m_{3/2}$; for instance, for $M_3=300\,$GeV and for a stau NLSP, the
maximal reheating temperature where $\Omega_{3/2}<1$, is $\sim
10^9\,$GeV. In the end it turns out that some fine-tuning betwen $M_3$,
$M_{\rm NLSP}$ and $m_{3/2}$ is required to find a region in which
$T_{\rm RH}$ can become as large as $\sim 10^9\,$GeV.

  There exist other potential renormalizable interaction terms that
violate messenger number by one unit. (Such operators lead to $1/m_{\rm Pl}^2$
suppressed proton decay \cite{DGP96}, due to gauge coupling of the messenger 
fields.) In the following we assume for
definiteness that messengers come in complete representations of
$SU(5)$, in particular, as $\mathbf{5}_M+\mathbf{\overline{5}}_M$ (or
$\mathbf{10}_M+\mathbf{\overline{10}}_M$), while the visible sector
superfields are denoted by $\mathbf{\overline{5}}_F+\mathbf{10}_F$,
and the Higgses sit in one pair of $\mathbf{5}_H+\mathbf{\overline
5}_H$ and one $\mathbf{24}_H$ supermultiplets.  Gauge symmetry limits
those interaction terms between messengers and visible sectors
superfields in the superpotential to the following:

\begin{eqnarray}
W_{\rm ren} & \supset \bigl\{ & \mathbf{\overline 5}_M\mathbf{\overline
5}_{F,H}\mathbf{10}_F\,,\, \mathbf{5}_M\mathbf{10}_F\mathbf{10}_F\,,\,
\mathbf{5}_M\mathbf{\overline 5}_{F,H}\mathbf{24}_H\,,\,\nonumber \\
&&
\mathbf{\overline 5}_M\mathbf{5}_{H}\mathbf{24}_H\,, \mathbf{\overline{10}}_M\mathbf{5}_H\mathbf{5}_H\,,\,
\mathbf{10}_M\mathbf{\overline{5}}_{H,F}\mathbf{\overline{5}}_{H,F}\,,\,\nonumber \\
&&
\mathbf{10}_M\mathbf{10}_F\mathbf{5}_H\,,\,
\mathbf{10}_F\mathbf{\overline{10}}_M\mathbf{24}_H
\bigr\}.
\label{eq:W-renorm}
\end{eqnarray} 

The interaction terms considered in Ref.~\cite{DNS97} are contained in
$\mathbf{\overline 5}_M\mathbf{\overline 5}_{H}\mathbf{10}_F$ and
$\mathbf{10}_M\mathbf{10}_F\mathbf{5}_H$, and lead to fast messenger
decay. Slow decay may, in principle, result from operators in
Eq.~(\ref{eq:W-renorm}) which involve particles with GUT scale masses.
However, renormalizable couplings to a $\mathbf{24}_H$ must be
excluded as they lead to $M_{\rm GUT}$ mass mixings between messengers
and visible sectors particles, hence they would spoil the
phenomenology at the electroweak scale.  The colored triplet GUT Higgs
in $\mathbf{5}_H$ and $\mathbf{\overline{5}}_H$ carry masses of order
$M_{\rm GUT}$ (but no {\it vev}), but they do not couple to the
lightest messenger, which is either the $5^{\rm th}$ component of a
$\mathbf{5}_M$ or the $(4,5)$ component of a $\mathbf{10}_M$. Hence
the colored Higgses do not lead to suppression of the decay width. In
Ref.~\cite{BM03} it was proposed that delayed messenger decay could
occur if the decay of the lightest messenger was suppressed by the
mediation of a particle of mass $\sim 10^{12}\,$GeV in a
renormalizable interaction. Our present discussion shows that this
model is not natural in the sense that it requires a new particle with
both the required mass $\sim 10^{12}\,$GeV and gauge charges such that
the required renormalizable interaction could occur. In the above list
of possible interactions terms, such coupling does not appear for the
minimal content of $SU(5)$.

   One may also argue that the required fine-tuning to avoid flavor
changing neutral currents (in particular for light messengers) may
actually indicate the absence of those renormalizable interactions,
and that messenger number violation occurs via further suppressed
interactions. Such couplings will be discussed further below.

\subsubsection{Renormalizable couplings in the K\"ahler function}

Fujii \& Yanagida~\cite{FY02} have proposed messenger-matter mixing
due to a correction in the superpotential $\delta W \simeq (\langle
W\rangle/m_{\rm Pl}^2){\mathbf 5}_M\overline{\mathbf 5}_F$, where
$\langle W\rangle \simeq m_{3/2}m_{\rm Pl}^2$. In the framework of
supergravity, a possible origin for such a superpotential term can be
highlighted by adding to the minimal K\"ahler potential $K_0
=\sum_i\Phi_i^{\dag}\Phi_i$, ($i$ running over all superfields
$\Phi_i$) a non-minimal part $\delta K$ given by,
\begin{equation}
\delta K = {\mathbf
5}_M\overline{\mathbf 5}_F + h.c
\label{eq:FY}.
\end{equation}
$\delta K$ is allowed by gauge
symmetries (and possibly by an R-symmetry as well, for conveniently
chosen R charges). Then, making use of the usual invariance of the
supergravity Lagrangian under K\"ahler transformations $K\to K +
F(\Phi )+F^*(\Phi^*)$, followed by superpotential $W\to e^{-F}W$ (and
super-Weyl) scalings, the above $\delta W$ is obtained for $F(\Phi ) =
- \delta K$ to the lowest order in $1/m_{\rm Pl}^2$, provided that a
constant is added to the superpotential to fine-tune the cosmological
constant after supersymmetry breaking (whence $\langle W\rangle \simeq
m_{3/2}m_{\rm Pl}^2$)\footnote{Altogether this is very reminiscent of
the Giudice-Masiero mechanism which provides a solution to the so-called 
$\mu$-problem \cite{MG}. }.  In our notations, the lightest messenger $\phi$ 
is a linear combination of the lightest scalar components of ${\mathbf
5}_M$ and $\overline{\mathbf 5}_M^*$, see Section~2.1. The mixing
between $\mathbf 5_M$ and $\overline{\mathbf 5}_F$ generated from
$\delta W \simeq m_{3/2}{\mathbf 5}_M\overline{\mathbf 5}_F$ thus
leads to the decay of $X$ into a lepton and a gaugino with width
$\Gamma_{X\rightarrow l\lambda} \simeq (g^2/16\pi)\,
m_{3/2}^2/M_X$~\cite{FY02}.

We should stress here that, starting as we do from $\delta K$ rather
than $\delta W$ of \cite{FY02}, one expects further 
contributions to the messenger decay or annihilation, with possibly important 
effects on the final gravitino abundance. Indeed, other contributions to the
decay into visible sector particles originate from the supergravity
scalar potential \cite{WBook}
\bea
V_B  =   e^{K/m_{\rm Pl}^2}&\Biggl[&K^{i\, j^{*}}\biggl(W \frac{K_i}{ m_{\rm Pl}^{2}}+W_i\biggr)
      \biggl(W^*\, \frac{K_{j^*}}{ m_{\rm Pl}^{2}}+W^*_{j^*}\biggr) 
\nonumber\\ && - \frac{3WW^*}{ m_{\rm Pl}^{2}}\biggr] \label{VB}
\eea 
\noindent
where, $i, j^{*}$ label the full set of scalar fields $\phi_i, \phi^{*}_{j^*}$; 
$K^{i\, j^{*}}$ denotes the inverse of the matrix 
$\partial K/\partial \phi_i\partial\phi^{*}_{j^*}$,
and $W_i = \partial W/\partial\phi_i$,
$W^*_{j^*} = \partial W^*/\partial\phi^*_{j^*}$.
From  $K \supset K_0 + \delta K$ and taking for illustration the case
of $\mathbf{5}+\mathbf{\overline{5}}$ messengers with
$ W \supset  S {\mathbf 5}_M\overline{\mathbf 5}_M +  y \overline{\mathbf 5}_F \overline{\mathbf 5}_H \mathbf{10}_F + 
\langle W\rangle$ one finds the leading contributions to the potential
 which induce the decay of the lightest messenger,

\bea V_B &\supset& m_{3/2}  S^* \mathbf{\overline
5}_F\mathbf{\overline 5}_M^* + y\, m_{3/2}\mathbf 5_M\mathbf{\overline
5}_H^*10_F^* -2 m_{3/2}^2 \mathbf 5_M \mathbf{\overline 5}_F \nonumber
\\ && + \frac{1}{m_{\rm Pl}^2} \biggl\{ \mathbf 5_M\mathbf{\overline
5}_F \Biggl[ |S|^2 ( \mathbf 5_M \mathbf 5_M^* + \mathbf{\overline
5}_M\mathbf{\overline 5}_M^* )   \nonumber \\ &&  \;\;\;\;\;\;\;\;\; +
\mathbf 5_M\mathbf{\overline 5}_M|^2 +y^2 |W_{y\vert i}|^2 \Biggr]
\nonumber \\ &&+ ( S^* \mathbf{\overline 5}_M^*
\mathbf{\overline 5}_F   + y \mathbf 5_M \mathbf{\overline 5}_H^*
\mathbf{10}_F^*) (S {\mathbf 5}_M\overline{\mathbf 5}_M +  
y \overline{\mathbf 5}_F \overline{\mathbf 5}_H \mathbf{10}_F)
\biggr\} \nonumber \\ && + h.c.
\label{VBFY}
\eea
\noindent
where we have neglected terms suppressed by higher powers of $1/m_{\rm
Pl}^2$ or of order $m_{3/2}/m_{\rm Pl}^2$ and smaller. Other operators
which do not induce messenger decay into standard model particles are
not shown, being irrelevant for the present discussion. After the
spurion scalar field $S$ has developed a supersymmetric {\it vev}, the
first term of Eq.~(\ref{VBFY}) leads to the bilinear operator
$m_{3/2}M_X\mathbf{\overline 5}_F\mathbf{\overline 5}_M^*$
contributing to the decay considered in \cite{FY02} (actually a
similar mixing between the fermionic partners is also generated, see
below). Note that the second operator in Eq.~(\ref{VBFY}) can
mediate an equally efficient decay of the lightest messenger to a
(colorless) Higgs and a slepton $\Gamma_{X\rightarrow H \tilde{f}}
\simeq (y^2/32\pi)\, m_{3/2}^2/M_X$, provided that the Yukawa coupling
is large (i.e. if $\delta K$ involves the third generation). A similar
decay to a Higgsino and a standard model fermion occurs as well,
triggered by the MSSM Yukawa vertex and $m_{3/2}M_X\mathbf{\overline
5}_F\mathbf{\overline 5}_M^*$. In any case, these new contributions do
not lead to a significant change in the analysis of the gravitino
relic density, being of the same order as $\Gamma_{X\rightarrow
l\lambda}$. 

Other contributions in Eq.~(\ref{VBFY}) can potentially lead to
important modifications when loop effects are considered. This is due
on one hand to the supersymmetry preserving non-renormalizable
operators of the form $\mathbf 5_M ... \phi \phi^*
... \mathbf{\overline 5}_F/m_{\rm Pl}^2$, with $\phi$ denoting an
arbritrary field, and on the other hand to renormalizable operators
which induce {\sl hard} supersymmetry-breaking after cancellation of
the cosmological constant.  (In connection with the latter operators,
the presence of the spurion, a visible sector singlet, could
destabilize the hierarchy of the messenger and/or electroweak scales
\cite{JBPR}.) A thorough discussion of these issues which are relevant
for the theoretical consistency of the effective supergravity model is
out of the scope of the present paper. Hereafter we give only a
partial and brief discussion of the two types of operators.\\ After
the {\sl vev} shift $ S \to S + M_X$, the term of order $m_{\rm
Pl}^{-2}$ in Eq.~(\ref{VBFY}) gives the operators \newline
$\biggl(\frac{M_X^2}{m_{\rm Pl}^2}\biggr) \times ( \mathbf
5_M\mathbf{\overline 5}_M\mathbf{\overline 5}_M^*\mathbf{\overline
5}_F , \mathbf 5_M\mathbf{\overline 5}_F \mathbf{\overline
5}_M\mathbf{\overline 5}_M^* , \mathbf 5_M\mathbf{\overline 5}_F
\mathbf 5_M \mathbf 5_M^*)$. These operators generate potentially very
large corrections to the $\mathbf 5_M-\mathbf{\overline 5}_F$ mixing,
through one-loop contributions of the $\mathbf 5_M$ and
$\mathbf{\overline 5}_M$, which are of the order of $\Lambda^2 \sim
m_{\rm Pl}^2$, where $\Lambda$ is the cut-off scale of the physics
underlying the effective supergravity Lagrangian. Even more, the
operators containing $S S^*$ instead of $M_X^2$ lead to more dangerous
corrections of order $\Lambda^4$ due to two-loop diagrams. The same
diagrams can also induce direct decays of the lightest messenger into
MSSM particles, such as $ X \to \gamma (Z) + \tilde{l}$. It is thus
important to assess the supersymmetric cancellations which would keep
those contributions under control, eventhough they originate from the
gravitational non-renormalizable sector.  The companion operators
involving the messenger fermions are contained in the $O(1/m_{\rm
Pl}^2)$ part of

\bea
- {\cal L}_F &\supset& e^{K/2m_{\rm Pl}^2} \biggl[ \frac{1}{2} {\cal D}_i D_j W
\bar{\chi}_R^i \chi_L^j + h.c. \biggr]
\label{LF1}
\eea
where $ {\cal D}_i D_j W = W_{ij} + (K_{ij}/m_{\rm Pl}^2) W + 
(K_i/m_{\rm Pl}^2) D_j W + (K_j/m_{\rm Pl}^2) D_i W -
(K_i K_j/m_{\rm Pl}^4)  W - \Gamma_{ij}^k D_k W$ 

\noindent
(see \cite{WBook}). They read 

\bea - {\cal L}_F \supset \frac{1}{2 m_{\rm Pl}^2} &\biggl[&
\overline{\psi}_R \psi_L \;  S \; (\mathbf
5_M\mathbf{\overline{5}}_F + h.c.)  \nonumber\\ && + \mathbf{\overline{\psi}}_R
\mathbf 5_M \; \mathbf{\psi}_L \mathbf{\overline{5}}_F \;  S
\biggr] + h.c.
\label{eq:LF}
\eea

\noindent
where $\psi$ denotes the Dirac field which combines the two fermionic
components of the $\mathbf{5}+\mathbf{\overline{5}}$ messenger
superfields. [The order of occurrence of the fields in
Eq.~(\ref{eq:LF}) indicates how they are combined into  $SU(5)$ invariants
.] In the supersymmetric limit, the one-loop contributions to the
$5_M-\bar{5}_F$ mixing induced by $- {\cal L}_F$ with $S \to \langle S
\rangle$ cancel exactly the ones from the scalar loops discussed
above. After SUSY breaking through the {\sl vev} of the auxiliary
field $F_S$, no dependence on the ultra-violet cut-off $\Lambda$ is
reintroduced. In particular, even $\log \Lambda$ terms cancel out
(though they would not have altered the size of the mixing) yielding a
correction of order $F_S^2/m_{\rm Pl}^2$ in the limit $F_S \ll M_X^2$,
and of order $M_X^4/m_{\rm Pl}^2$ in the limit $F_S \simeq M_X^2$,
which remains negligible when compared to the tree-level mixing
magnitude $m_{3/2} M_X$, Eq.~(\ref{VBFY}), in the parameter space
region relevant for gravitino dark matter.\footnote{We checked also
for cancellations to two-loop order considering subclasses of Feynman
diagrams which involve the spurion scalar and
fermion virtual contributions. A detailed study is outside the scope
of the present paper.} Finally, as mentioned before, some {\sl hard}
SUSY breaking operators are generated in Eq.~(\ref{LF1}) from $m_{3/2}
e^{K/2m_{\rm Pl}^2} ( \frac{1}{2} \bar{\chi}_R^i \chi_L^j \delta
K_{ij}+ h.c.)$, leading to bilinear matter fermion mixing between the
messengers and the MSSM particles, $m_{3/2} (\overline{f}_R \psi_L +
h.c.)$. These can potentially lead to quadratic divergences which
would destabilise the scale of the mixing $\mathbf{5}_M -
\mathbf{\overline 5}_F$ between the lightest messenger and the MSSM
scalar particles. However, leading  one-loop (tadpole) effects with one
mass insertion occur as corrections to $\mathbf{\overline 5}_M^* -
\mathbf{\overline 5}_F$ and turn out to be at worse
$O((\Lambda^2/m_{\rm Pl}^2) m_{3/2} M_X)$, thus harmless for a cut-off
of order $\sim m_{\rm Pl}$. (Note also that the matter fermion kinetic terms
induce fermionic tadpole contributions which cancel among themselves
due to the derivative couplings.)  To summarize, in the sector of the
supergravity Lagrangian not involving the gravitino the $\delta K$
piece of the K\"ahler potential leads essentially to the same features
as discussed in \cite{FY02}, at least to leading order in $1/m_{\rm
Pl}^2$ and up to one-loop. One exception is the messenger decay induced by
first term in Eq.~(\ref{VBFY}) which we will consider later on.  
 
\begin{figure*}
  \centering
  \includegraphics[width=0.9\textwidth,clip=true]{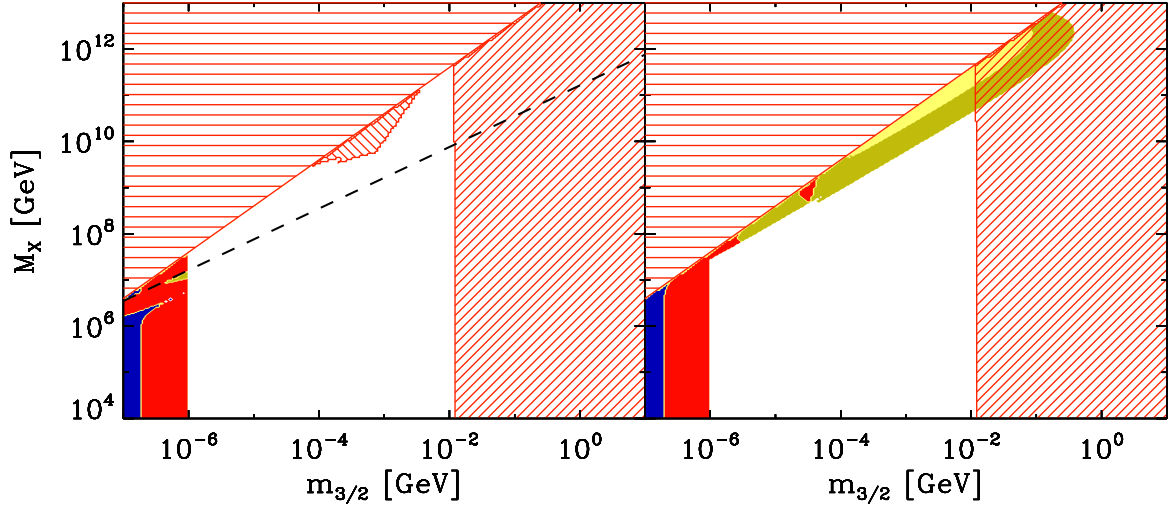}
  \caption[...]{Contours of $\Omega_{3/2}$ in the plane $M_X-m_{3/2}$
  for one pair of messengers sitting in
  $\mathbf{5}+\mathbf{\overline{5}}$ representations; the lightest
  messenger $X$ carries the gauge charges of a $\tilde\nu_L$, and
  decays into a slepton and a gaugino via the mixing term $\mathbf
  5_M\mathbf{\overline{5}_F}$ in the K\"ahler function. Color shading
  is as in Fig.~\ref{fig:fast-T}; the NLSP is assumed to be a bino in
  both panels. In the left panel, the spurion $S$ is assumed to be
  much heavier than the lightest messenger and annihilation
  $XX^*\rightarrow\tilde G\tilde G$ proceeds with the cross-section of
  Eq.~\ref{eq:ann-gold-lo}; the dashed line indicates the limit above which
  this cross-section  violates the unitarity bound, see
  Section~\ref{sec:2}. In the right panel, the spurion $S$ is assumed
  much lighter than $X$, and annihilation into goldstinos is given by
  Eq.~\ref{eq:ann-gold-hi}. See text for further details.}
\label{fig:FY}
\end{figure*}

  The relic abundance for the gravitino can be calculated using the
techniques developed in Section~III and the results are shown in
Fig.~\ref{fig:FY}. This figure uses the color shading as in
Fig.~\ref{fig:fast-T} but is plotted in the plane $m_{3/2}-M_X$
instead of $T_{\rm RH}$, which was taken to be $T_{\rm
RH}=10^{12}\,$GeV.  The plot shown in the left panel assumes that the
spurion $S$ is much heavier than the lightest messenger $X$, in which
case annihilation $XX^*\rightarrow \tilde G\tilde G$ takes place with
the cross-section given in Eq.~(\ref{eq:ann-gold-lo}). As discussed in
Section~II.B, this cross-section violates unitarity for $M_X\gtrsim
10^7\,{\rm GeV}(m_{3/2}/1\,{\rm keV})^{2/3}$, i.e. in the region above
the dashed thick line in the left panel of Fig.~\ref{fig:FY}. There is
no solution in this case for gravitino dark matter, at variance with
the conclusions of Ref.~\cite{FY02}. The main reason is that
annihilation into goldstinos has not been accounted for in
Ref.~\cite{FY02}, yet the solution for gravitino dark matter proposed
by these authors lies in the region in which unitarity is
violated. The annihilation cross-section at its unitarity bound is
much larger than that used in Ref.~\cite{FY02} for annihilation
through gauge interactions, hence the messenger relic abundance and
the amount of gravitino dilution are correspondingly smaller. At the
very least, since one cannot predict the cross-section in this region
where multi-goldstino production violates unitarity, one can conclude
that the results for the scenario of Ref.~\cite{FY02} in this region
are model dependent in that they require contributions from the hidden
sector to bring down multi-goldstino production to a negligible level.

  In the right panel of Fig.~\ref{fig:FY}, it is assumed that $S$ is
much lighter than $X$. Annihilation into goldstinos scales with the
effective Yukawa coupling $F_S/F$ of the messenger to the goldstino
component of the fermionic partner of $S$, as in
Eq.~(\ref{eq:ann-gold-hi}). This annihilation channel thus contributes
only in the region of direct GMSB scenarios where $k\equiv F_S/F \sim
1$, which happens to be that where gravitino dark matter can be found
for the mixing term $\mathbf{5}_M\overline{\mathbf{5}}_F$ proposed in
Ref.~\cite{FY02}.  The lightest messenger can also annihilate into a
pair of spurions, as discussed in Section~II.B.  Furthermore, one must
also consider the possible decay of the lightest messenger into one
visible sector sneutrino and two goldstinos, which is induced by the
above mixing term and $XX^*\tilde G\tilde G$ four-point vertices
similar to the ones discussed after Eq.~(\ref{eq:ann}). For instance,
the latter vertex is induced by terms of the form $ -e^{K/2m_{\rm
Pl}^2} W \bar{\Psi}_L^\mu \sigma_{\mu \nu} \Psi_R^\nu \; /m^2_{\rm
Pl}$, \cite{WBook}, using Eqs.~(\ref{eq:FY}, \ref{eq:corresp1}) and $
W \to \langle W\rangle$.

One expects this decay width to scale as
$\Gamma_{X\rightarrow\tilde\nu\tilde G\tilde G}\sim 0.1
(F_S/F)^4\Gamma_{X\rightarrow l\lambda}$, with the prefactor of $0.1$
accounting for the enlargement of phase space.  This decay channel
produces highly relativistic gravitinos which mix with the ``cold''
gravitinos produced by other channels, see Section~III.A, resulting in
mixed dark matter. The small area around $m_{3/2} \sim 10-100\,$keV is
shown in red (medium shading) in Fig.~\ref{fig:FY}, indicating that
the hot gravitinos contribute to more than 10\% of the gravitino
energy density and that the averaged velocity corresponds to warm dark
matter. Finally one must also include decay into a spurion and a
sneutrino induced by the three scalar coupling, first term in
Eq.~(\ref{VBFY}), with width $\Gamma_{X\rightarrow S\tilde\nu}\simeq
(1/16\pi)m_{3/2}^2/M_X$.

The inclusion of these various annihilation and decay channels modify
the region of parameter space where one can find gravitino dark matter
with respect to the conclusions of Ref.~\cite{FY02}. The various
effects add up in pushing this region to higher values of
$M_X$. In effect, the increased lightest messenger annihilation cross-section 
decreases their relic abundance hence the amount of entropy production;
similarly, the increased lightest messenger decay width increases their decay 
temperature hence also decreases the entropy production. These effects can be
compensated, at a given value of $m_{3/2}$, by increasing $M_X$
since $\langle \sigma_{XX^*\rightarrow\ldots} v\rangle \propto 1/M_X^2$ and
$\Gamma_{X\rightarrow\ldots}\propto 1/M_X$.

  Note that Fig.~\ref{fig:FY} assumes that the NLSP is a bino; as seen
previously, the BBN bounds would be relaxed is the NLSP turned out to
be a stau.

 Finally, the influence of the choice of the post-inflationary
reheating temperature (taken as $T_{\rm RH}=10^{12}\,$GeV here) is as
follows. For this chosen value of $T_{\rm RH}$, gravitinos (i.e. their
1/2 component) abundances are brought to equilibrium for all
$m_{3/2}\lesssim 100\,$MeV.  For smaller values of $T_{\rm RH}$, the
plot would thus look similar in the region where $M_X\lesssim T_{\rm
RH}$, unless $T_{\rm RH}$ is that low $T_{\rm RH}\simle 10^{12}{\rm
GeV} (m_{3/2}/0.1\,{\rm GeV})^2$ that gravitinos are initially not in
chemical equilibrium. For $M_X\gtrsim T_{\rm RH}$, messengers are not
produced at the post-inflationary reheating, there is no entropy
production (gravitino dilution) and the gravitino abundance in the
plane $m_{3/2}-T_{\rm RH}$ then resembles that shown in the right
panels of Fig.~\ref{fig:fast-T}.

If the messengers sit in $\mathbf{10}+\mathbf{\overline{10}}$
representations of $SU(5)$, a similar mixing term
$\mathbf{\overline{10}}_M\mathbf{10}_F$ can be induced, and leads to
similar effects. Note that the mixing ${\mathbf 5}_M\overline{\mathbf
5}_F$ and $\mathbf{\overline{10}}_M\mathbf{10}_F$ are constrained by
$R-$symmetry charge assignments. If a non-holomorphic mixing
$\overline{\mathbf 5}_M^*\overline{\mathbf 5}_F$ (or $\mathbf
{\overline{10}}_M^*\mathbf{\overline{10}}_F$) is allowed, the
situation is quite different. In effect, the lightest messenger now
mixes in the kinetic terms with the sneutrino and its decay is no
longer suppressed by $m_{3/2}/M_X$. As a consequence, decay is much
faster, $\Gamma\sim{\cal O}(M_X)$, and no entropy production occurs;
the situation is then as shown in Fig.~\ref{fig:fast-T}.

\subsection{Non-renormalizable couplings}

\subsubsection{Superpotential interactions}

It is possible that interactions of the type Eq.~(\ref{WDine}) are
forbidden and that messengers may only decay via non-renormalizable
operators in $W$, as discussed previously. Operators which may arise
due to unknown Planck-scale physics, which respect the $SU(5)$ gauge
symmetry and that violate the messenger number by one unit are given
to leading order in $1/m_{\rm Pl}$, by:

\begin{eqnarray}
 W_{\rm non-ren} & \supset \frac{1}{m_{\rm Pl}}\,\bigl\{ & \mathbf{\overline 5}_M \mathbf
{10}_F\mathbf{10}_F\mathbf{10}_F\,,\, {\mathbf 5}_M{\mathbf
5}_{H}\mathbf{\overline 5}_{H,F}\mathbf{\overline 5}_{H,F}\,,\,\nonumber\\
& &
\mathbf{\overline{5}}_M{\mathbf 5}_{H}\mathbf{5}_{H}\mathbf{\overline
5}_{H,F}\,,\, {\mathbf 5}_M{\mathbf 5}_{H}{\mathbf
5}_{H}\mathbf{10}_F\,,\, \nonumber\\
& & \mathbf{\overline 5}_M\mathbf{\overline 5}_H\mathbf{10}_F\mathbf{24}_H\,,\,
\mathbf{5}_M\mathbf{\overline 5}_{H,F}\mathbf{24}_H\mathbf{24}_H\,,\,\nonumber\\
& &
\mathbf{\overline 5}_M\mathbf{5}_H\mathbf{24}_H\mathbf{24}_H\,,\,
\mathbf{10}_F\mathbf{\overline{10}}_M 
{\mathbf 5}_{H}\mathbf{\overline 5}_{H,F}\,,\,\nonumber\\
& & 
\mathbf{\overline{10}}_M\mathbf{\overline{5}}_{H,F}\mathbf{\overline{5}}_{H,F}\mathbf{\overline{5}}_{H,F}\,,\,
\mathbf{10}_M\mathbf{5}_{H}\mathbf{5}_{H}\mathbf{5}_{H}\,,\,\nonumber\\
& &
\mathbf{10}_M\mathbf{\overline{5}}_{H,F}\mathbf{10}_F\mathbf{10}_F\,,\,
\mathbf{10}_M\mathbf{10}_F\mathbf{5}_H\mathbf{24}_H\,,\,\nonumber\\
& & \mathbf{\overline{10}}_M\mathbf{5}_{H}\mathbf{5}_{H}\mathbf{24}_H\,,\,
\mathbf{10}_M\mathbf{\overline{5}}_{H,F}\mathbf{\overline{5}}_{H,F}\mathbf{24}_H\,,\,\nonumber\\
& & 
\mathbf{\overline{10}}_M\mathbf{10}_F\mathbf{24}_H\mathbf{24}_H\,, \,\mathbf{5}_M\mathbf{\overline{5}}_M\mathbf{5}_M\mathbf{\overline{5}}_F\,,\,\nonumber\\
& & 
\mathbf{10}_M\mathbf{\overline{10}}_M\mathbf{10}_M\mathbf{10}_F
\bigr\}
\label{eq:W-nonrenorm} 
\end{eqnarray}

\noindent
All terms in Eq.~(\ref{eq:W-nonrenorm}) which involve couplings of one
lightest messenger to $\mathbf{\overline 5}_F$ or $\mathbf{10}_F$ but
not Higgses lead to decay into three-body final states with decay
width $\sim 10^{-4} M_X^3/m_{\rm Pl}^2$ . It is easy to see, using
Eq.~(\ref{Delta2}) that entropy production is not sufficient to dilute
the gravitinos to the required abundance for a high post-inflationary
reheating temperature $T_{\rm RH}\gg 10^8\,$GeV (and $T_{\rm RH}\gg
M_X$). Admittedly this is a drawback of the present scenario since
those terms are the most generic.

  Let us now consider the terms involving couplings to Higgses. For
terms involving one $\mathbf{24}_H$ acquiring {\it vevs} of order $\sim
M_{\rm GUT}$, say $X\Phi_1\Phi_2 H_{\mathbf 24}/m_{\rm Pl}$, the
Lagrangian contains the effective Yukawa interaction $\sim (M_{\rm
GUT}/m_{\rm Pl})X\overline\chi_1\chi_2$ from the fermionic part of the
Lagrangian contained in Eq.~(\ref{LF1}), with $\chi_1$ and $\chi_2$
the Weyl spinors of $\Phi_1$ and $\Phi_2$. This Yukawa interaction
between the lightest messenger and two fermions with effective
coupling constant $\sim 10^{-3}$ leads to fast decay if $\chi_1$ and
$\chi_2$ have electroweak scale masses. Other terms in the Lagrangian
lead to similar or only somewhat smaller partial widths. Consequently
the conclusions of the previous section with regards to gravitino dark
matter with "fast" decaying messengers apply. From
Eq.~(\ref{eq:W-nonrenorm}) one can check that all possible
above combinations involving one $\mathbf{24}_H$ contain a coupling of
$X$ to particles with electroweak masses except
$\mathbf{10}_M\mathbf{10}_F\mathbf{5}_H\mathbf{24}_H$. However, even
for the latter term, the scalar potential contains the interaction
$\mathbf{10}_M\mathbf{10}_F\mathbf{24}_Hy\mathbf{10}_F^*
\mathbf{10}_F^*/m_{\rm Pl}$ generated by $\vert\partial W/\partial
\mathbf{5}_H\vert^2$, with $y$ the third family Yukawa coupling. This
interaction gives a decay width $\Gamma \sim 10^{-5} y^2(M_{\rm
GUT}/m_{\rm Pl})^2 M_X$, which is too large to allow solutions for
gravitino dark matter.

  Terms involving two $\mathbf{24}$ should be excluded as they lead to
unacceptable mass mixings between messengers and visible sectors
superfields.

  Consider now terms of the form $X\Phi_1\Phi_2\Phi_3/m_{\rm Pl}$
containing at least one $\mathbf{5}_H$ but no $\mathbf{24}_H$. It can
be checked that all terms of this form in Eq.~(\ref{eq:W-nonrenorm}) 
contain couplings of $X$ to particles with electroweak masses,
hence lead to fast decay as above, except for
$\mathbf{5}_M\mathbf{5}_H\mathbf{5}_H\mathbf{10}_F/m_{\rm Pl}$,
$\mathbf{10}_M\mathbf{5}_H\mathbf{5}_H\mathbf{5}_H/m_{\rm Pl}$ and
$\mathbf{\overline{10}}_M\mathbf{\overline 5}_{H,F}\mathbf{\overline
5}_{H,F}\mathbf{\overline 5}_{H,F}/m_{\rm Pl}$.  The first of these
terms, when written for the lightest messenger, contains at least one
colored Higgs, say $H_i$. However, the scalar potential term
$\vert\partial W/\partial \mathbf{5}_{H_i}\vert^2$ generates here as
well an interaction between the lightest messenger with 4 particles of
electroweak masses, leading to decay width $\Gamma \sim 10^{-5}y^2
M_X^3/m_{\rm Pl}^2$, still too large for dark matter solutions.
The term $\mathbf{10}_M\mathbf{5}_H\mathbf{5}_H\mathbf{5}_H/m_{\rm
Pl}$ couples $X$ to the three colored Higgses, hence its decay is too
highly suppressed both by the GUT scale and phase space, $\Gamma \sim
10^{-12} (M_X/M_{\rm GUT})^8M_X^3/m_{\rm Pl}^2$, and cannot lead to
decay before BBN for $M_X\lesssim 10^{14}\,$GeV. Note that the
big-bang nucleosynthesis constraints on NLSP decay can be translated
into an extreme upper bound $M_X \lesssim 10^{12}\,$GeV (see
Fig.~\ref{fig:FY} and \cite{GGR99}).  Finally, for the term
$\mathbf{\overline{10}}_M\mathbf{\overline 5}_{H,F}\mathbf{\overline
5}_{H,F}\mathbf{\overline 5}_{H,F}/m_{\rm Pl}$, similar conclusions
apply if three Higgses are involved.  For couplings involving two
colored Higgses, decay is also too highly suppressed.  In effect, the
lightest messenger can then decay into 5-body final state with
mediation by a GUT mass Higgs, leading to $\Gamma \sim
10^{-6}y^2(M_X/M_{\rm GUT})^4M_X^3/m_{\rm Pl}^2$.
It cannot decay before BBN if $M_X \lesssim 10^{12}\,$GeV, as can be
checked using Eq.~(\ref{eq:Tda}). Finally, if only one Higgs is
involved in the coupling, the messenger can decay into four particle
final states with decay width $\Gamma \sim 10^{-5}y^2 M_X^3/m_{\rm
Pl}^2$ and yet provide no solution for gravitino dark matter.

The last two operators of Eq.~(\ref{eq:W-nonrenorm}) together with the
first term of Eq~(\ref{eq:W1}) induce at one-loop order a mixing
between the lightest messenger and the MSSM matter fields.  This
mixing is non-vanishing only after supersymmetry breaking and is found
to be of magnitude $ \sim (\sqrt{2}M_X/m_{\rm Pl})\, F_S\, \log
(\Lambda/M_X)$. As discussed  in Section~IV.A.2, $\Lambda$ is
identified with a physical cut-off of order $m_{\rm Pl}$, however, in
contrast with the results of that section, 
there is here no (accidental) cancellation of the cut-off dependence and no
suppression by the gravitino mass. The mixing is so large that decay is not 
accompanied by entropy
production, and consequently there is no solution for gravitino dark
matter.

  In summary, non-renormalizable interaction terms in the
superpotential for the minimal content of GMSB scenarios in $SU(5)$
grand unification do not allow for natural solutions leading to
gravitino dark matter (unless the post-inflationary reheating
temperature is tuned as before).

\subsubsection{Non-renormalizable interactions in the K\"ahler function}

Interaction terms in the K\"ahler potential $K$ can either be
holomorphic or not in the superfields, leading to different
phenomenologies, which we explore in turn.  To leading order in
$1/m_{\rm Pl}$, non-renormalizable holomorphic operators have the same
form as those shown in Eq.~(\ref{eq:W-renorm}),
\begin{equation}
K_{\rm hol} = \frac{W_{\rm ren}}{m_{\rm Pl}} + h.c.
\end{equation}
 

\noindent
We write generically these operators
 as $X\phi_1\phi_2/m_{\rm Pl}$. Let us assume for the
moment that both $\phi_1$ and $\phi_2$ have masses of order of the
electroweak scale.

  The K\"ahler $U(1)$ connection $K_j\partial_\mu\phi_j - h.c.$, with
$K_j\equiv \partial K /\partial \phi_j$ induces the following coupling
to the fermionic components $\psi^i$ of all the chiral superfields of
the model ~\cite{WBook}, 
\begin{equation}
- {\cal L}_F \supset {1\over 2 m_{\rm
Pl}^3}  \overline{\psi}_L^i\gamma^\mu\psi^i_L \; {\rm Im} \; \partial_\mu(X\phi_1\phi_2),
\label{Uconn}
\end{equation}
 which leads to a partial decay width $\propto M_X^7/m_{\rm Pl}^6$. From
Eq.~(\ref{LF1}) an
effective Yukawa coupling is generated, 
\begin{equation}
{\partial^2K\over\partial\phi_1\partial\phi_2}{\langle W \rangle \over
m_{\rm Pl}^2}\overline{\psi}^1_R\psi^2_L\sim {m_{3/2}\over m_{\rm
Pl}}X\overline{\psi}^1_R\psi^2_L,\end{equation} which leads to a
highly suppressed partial decay width $\sim (m_{3/2}/m_{\rm
Pl})^2M_X^3/m_{\rm Pl}^2$.
 Finally, couplings to goldstinos are generated notably by the
gravitino mass term and gravitino kinetic terms:
\begin{equation}
{-1\over 2m_{\rm Pl}} X\phi_1\phi_2 {W\over m_{\rm
Pl}^2}\Psi_\mu\sigma^{\mu\nu}\Psi_\nu + \epsilon^{\mu\nu\rho\sigma}
{1\over m_{\rm Pl}^{3}}\partial_\rho
(X\phi_1\phi_2)\overline\Psi_\mu\overline\sigma_\nu\Psi_\sigma.
\end{equation}

\noindent 
with the replacement $\Psi_\mu \to i \sqrt{2/3}\;\partial_\mu G
/m_{3/2}$. 

\begin{figure*}
  \centering
  \includegraphics[width=0.9\textwidth,clip=true]{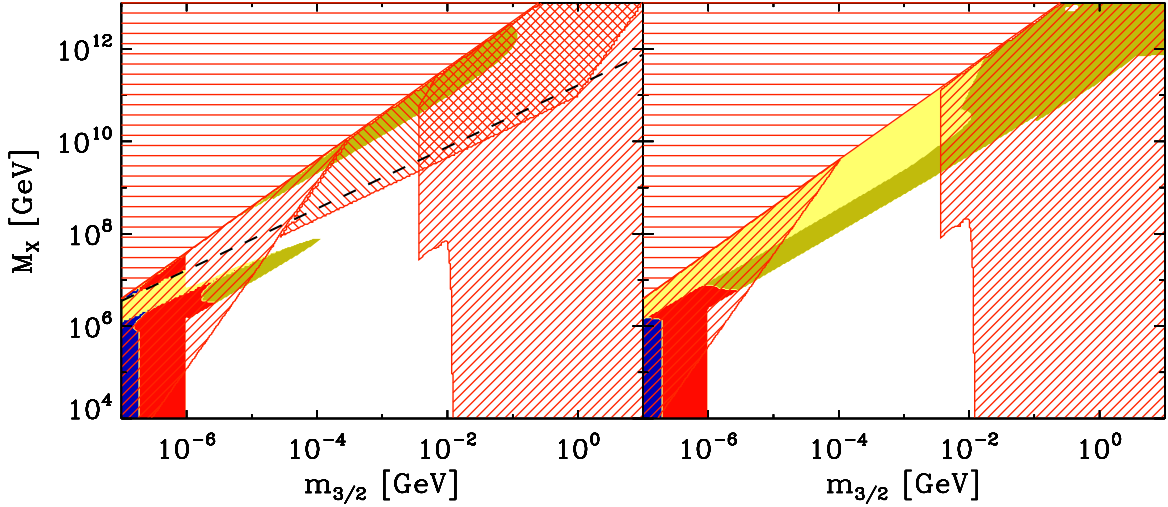}
  \caption[...]{Contours of $\Omega_{3/2}$ in the plane $M_X-m_{3/2}$
  for one pair of messengers sitting in
  $\mathbf{5}+\mathbf{\overline{5}}$ representations; the lightest
  messenger $X$ decays into two goldstinos and one sfermion or one
  sfermion and one gaugino via the mixing term $\sim X \phi_1(M_{\rm
  GUT}/m_{\rm Pl})$ in the K\"ahler function. Same color code and
  conventions as in
  Figs.~\ref{fig:fast-T},~\ref{fig:FY}. The red NE-SW dashed area in the left 
  part of the figure results from big-bang nucleosynthesis constraints on 
  lightest messenger decay and the NW-SE area is forbidden by the contribution
  of gravitinos to the energy density at BBN; see text for further
  details.}
\label{fig:Khol-24}
\end{figure*}

 All terms lead to extremely slow decay if $\phi_1$ and $\phi_2$ have
electroweak scale masses, and must be forbidden in order for the
lightest messenger not to decay after BBN.  However if one
$\mathbf{24}_H$ is present, the replacement of this field by its {\it
vev} in the K\"ahler function shows that one recovers a mixing term as
proposed in Ref.~\cite{FY02} albeit with effective coupling $M_{\rm
GUT}/m_{\rm Pl}$. This mixing term then leads to the same decay widths
into one sfermion and one gaugino, or one sfermion and two goldstinos,
as discussed before, albeit decreased by $(M_{\rm GUT}/m_{\rm
Pl})^2\sim 10^{-5}$. The consequences for gravitino dark matter are
shown in Fig.~\ref{fig:Khol-24}. As discussed before, we should
consider the cases where the spurion is heavier or lighter than the
lightest messenger.  The left panel shows the case where the spurion
is heavier than the lightest messenger and the annihilation
cross-section into goldstinos increases with increasing $M_X$ to
saturate at the unitarity bound above the dashed line.  Only a small
portion of parameter space allows for gravitino dark matter in this
case; it is actually an amputated part of the solution shown in the
right panel, see below. Above the dashed line, the results are highly
uncertain since multi-goldstino is not well controled. Moreover, the
lightest messenger can decay into a sneutrino and a pair of goldstinos
due to the above mixing, with a decay width which is expected to scale
as $\Gamma_{X\rightarrow \tilde\nu\tilde G\tilde G}\propto
(F_S^2M_X^4/F^4)\Gamma_{X\rightarrow l\lambda}$. The prefactor
$(F_S^2M_X^4/F^4)$ denotes the effective coupling of $XX^*$ to a pair
of goldstinos in the heavy spurion limit. In the region in which
unitarity is violated, it has been assumed that this effective
coupling saturates at the value reached [$\sim {\cal O}(1)$] when the
annihilation cross-section reaches the unitarity bound. It is then
comparable to the decay width into a lepton and a gaugino and produces
highly relativistic gravitinos. The energy density contained in these
gravitinos exceeds the BBN bounds on additional relativistic degrees
of freedom so that most of this region is excluded, as indicated by
the NW-SE oriented dashed lines in Fig.~\ref{fig:Khol-24}. The SW-NE
oriented dashed lines exclude the part of parameter space at small
$m_{3/2}$ in which the lightest messenger decay occurs so late that it
is forbidden by BBN constraints on energy injection. If the NLSP were
a stau, this region would still be forbidden but the constraints at
large $m_{3/2}$ would be relaxed.

In the right panel of Fig.~\ref{fig:Khol-24}, the spurion is assumed
to be lighter than the lightest messenger and annihilation into
goldstinos is less effective. One finds a solution for gravitino dark
matter in a large part of parameter space, $m_{3/2}\sim 10\,{\rm
keV}\rightarrow 1\,{\rm MeV}$, for scenarios of indirect gauge
mediation, i.e. $k\ll1$. This solution is the same as  that
discussed in Section~IV.A.2, albeit
shifted to smaller values of $M_X$; this can be understood from the
fact that for a same relic abundance of $X$, entropy production is
larger in the present case since the decay width is further
suppressed. Hence one can tolerate a smaller relic abundance, or
equivalently a higher annihilation cross-section, i.e. a smaller $M_X$.

Finally consider now non-holomorphic non-renormalizable couplings
between $X$ and visible sector particles in $K$. Such couplings can
take the form:

\begin{eqnarray}
K & \supset \, {1\over m_{\rm Pl}}\bigl\{&
\mathbf{5}_M^\dagger\mathbf{\overline{5}}_{H,F}\mathbf{10}_F\,,\,
\mathbf{\overline{5}}_M\mathbf{5}_H^\dagger\mathbf{10}_F\,,\,
\mathbf{\overline{5}}_M^\dagger\mathbf{10}_F\mathbf{10}_F\,,\,\nonumber\\
& &
\mathbf{5}_M^\dagger\mathbf{5}_{H}\mathbf{24}_H\,,\,
\mathbf{5}_M\mathbf{5}_{H}^\dagger\mathbf{24}_H\,,\,
\mathbf{\overline{5}}_M\mathbf{\overline{5}}_{H,F}^\dagger\mathbf{24}_H\,,\,\nonumber\\
& &
\mathbf{\overline{5}}_M^\dagger\mathbf{\overline{5}}_{H,F}\mathbf{24}_H\,,\,
\mathbf{10}_M^\dagger\mathbf{5}_H\mathbf{5}_H\,,\,
\mathbf{\overline{10}}_M\mathbf{\overline{5}}_{H,F}^\dagger\mathbf{5}_H\,,\,\nonumber\\
& &
\mathbf{\overline{10}}_M^\dagger\mathbf{\overline{5}}_{H,F}\mathbf{\overline{5}}_{H,F}\,,\,
\mathbf{\overline{10}}_M^\dagger\mathbf{10}_F\mathbf{5}_H\,,\,
\mathbf{10}_M\mathbf{10}_F\mathbf{\overline{5}}_{H,F}^\dagger\,,\,\nonumber\\
& &
\mathbf{10}_M^\dagger\mathbf{10}_F\mathbf{24}_H\,,\, 
\mathbf{10}_M\mathbf{10}_F^\dagger\mathbf{24}_H\,,\,+\, {\rm
h.c.}\bigr\}
\label{eq:K-nonrenorm}
\end{eqnarray}

As before, we write this coupling as $X^*\phi_1\phi_2/m_{\rm Pl}$ and
assume for the moment that $\phi_1$ and $\phi_2$ carry electroweak
scale masses. Then decay into $\phi_1$ and $\phi_2$ with width $\Gamma
\sim 10^{-2}M_X^3/m_{\rm Pl}^2$ occurs via the mixing of kinetic terms
between $\phi_1$ and $X$ or between $\phi_2$ and $X$. As seen before,
such a decay width does not lead to solutions for gravitino dark
matter as entropy production is not significant.  Inspection of
Eq.~(\ref{eq:K-nonrenorm}) reveals that all terms fall in the above
category except those involving one $\mathbf{24}_H$ as well as
$\mathbf{\overline{10}}_M^\dagger\mathbf{10}_F\mathbf{5}_H$ and
$\mathbf{10}_M\mathbf{10}_F\mathbf{\overline{5}}_{H}^\dagger$.

  The latter two terms necessarily contain one colored Higgs with GUT
mass but no {\it vev}, which we assume to be $\phi_2$. Then the scalar
potential term involving the inverse of the K\"ahler metric $g$
contains a coupling of $X$ to visible sector particles:
$g^{\phi_2\,M^*}D_{\phi_2}WD_{M^*}W^* \supset (\phi_1/m_{\rm Pl})y
M_X\mathbf{10}_F\mathbf{10}_F X$. This leads to decay into three-body
final state with width $\Gamma \sim 10^{-4}y^2 M_X^3/m_{\rm Pl}^2$,
again too large to yield solutions for gravitino dark matter.

  Finally, if coupling to one $\mathbf{24}_H$ occurs, say $\phi_2$,
mass mixing of order $M_{\rm GUT}M_X$ occurs between $\phi_1$ and $X$
and leads to one negative mass squared eigenstate; this coupling must
therefore be excluded.

\section{Discussion}\label{sec:5}

\begin{figure*} \centering
  \includegraphics[width=0.9\textwidth,clip=true]{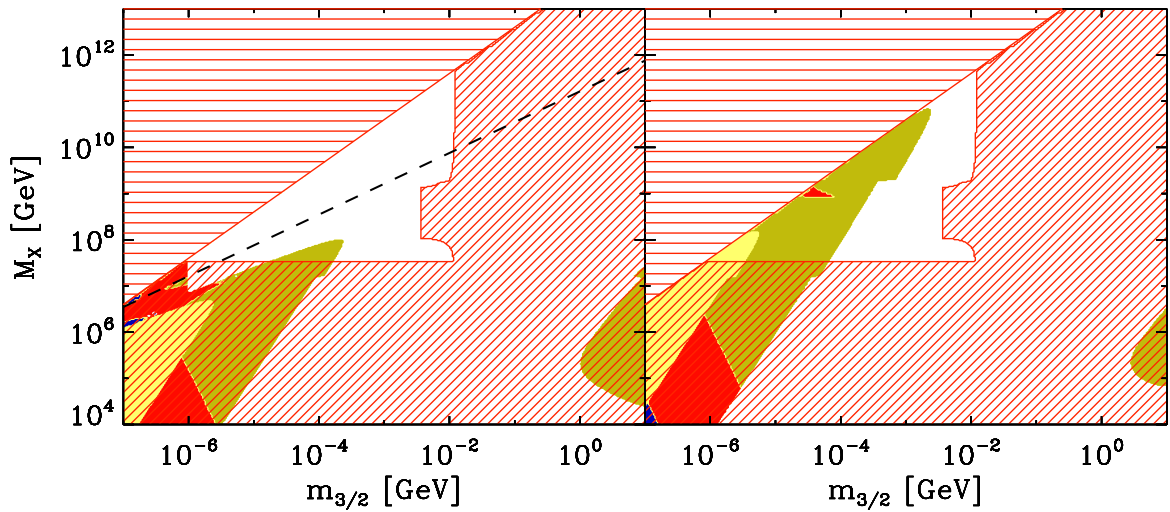}
  \caption[...]{Contours of $\Omega_{3/2}$ in the plane $M_X-m_{3/2}$
  for one pair of messengers sitting in
  $\mathbf{5}+\mathbf{\overline{5}}$ representations as in
  Fig.~\ref{fig:FY}, but for decay width $\Gamma \sim
  10^{-10}M_X^3/m_{\rm Pl}^2$. See text for details.}
\label{fig:555524}
\end{figure*}

 In this section we explore qualitatively other possible avenues which
may help reconcile gravitino dark matter with more generic GMSB
scenarios.  Indeed, the previous discussion has shown that interesting
solutions for gravitino dark matter and/or the gravitino problem in
GMSB with $SU(5)$ grand unification with a high post-inflationary
reheating temperature, can only be found for some very specific
couplings between the messenger and visible sector and in some
restricted regions of parameter space. It is furthermore
necessary to assume that the spurion is much lighter than the 
lightest messenger so that multi-goldstino production remains at a
safe level. A gravitational decay width 
$\Gamma \sim 10^{-3} M_X^3/m_{\rm Pl}^2$, which is generic
in the sense that it is generated by most allowed non-renormalizable
messenger-matter interaction terms, does not lead to satisfying
solutions for gravitino dark matter. 

  It is instructive to consider the case of decay widths with the same
scaling but whose prefactor is much smaller, $\Gamma \sim \epsilon
M_X^3/m_{\rm Pl}^2$ with $\epsilon\ll 10^{-3}$.
Figure~\ref{fig:555524} shows the solution for $\epsilon=10^{-10}$.
Such decay width can be achieved by terms of the form $W\supset
X\Phi_1\Phi_2\Phi_3 H_{24}/m_{\rm Pl}^2$, with $H_{24}$ designing a
Higgs with non-zero {\it vev} in $\mathbf{24}_H$, or by most
non-renormalizable couplings to order $1/m_{\rm Pl}$ discussed in the
previous section, provided they are further suppressed by a factor
$\sim 10^{-7}$.  In this figure, one finds that in the left panel,
where $S$ is assumed heavier than $X$, i.e. where multi-goldstino
production plays a significant role, there is room for gravitino dark
matter only in a very limited region of parameter space. In this area,
furthermore, the post-decay reheating temperature is quite close to
$1\,$MeV. On the contrary, in the right panel one recovers a solution
for gravitino dark matter with mass $m_{3/2}\sim 10\,{\rm
keV}\,\rightarrow 10\,{\rm MeV}$ in direct gauge mediated scenarios
$k\lesssim 1$.  Higher values of the coupling $\epsilon$ lead to
solutions shifted to higher $M_X$, with the dashed region due to late
messenger decay shifting downwards. Lower values of $\epsilon$ lead to
solutions shifted to smaller values of $M_X$, but with the excluded
dashed region moving upwards in $M_X$.

  Dangerous operators involving GUT-scale {\sl vev}'s may also be
present.  However, global continuous R-symmetries are expected to play
an important r\^ole in scenarios of supersymmetry breaking
\cite{BKN94}, and in a generic setting they could control the absence
of unwanted operators for properly chosen R-charges
\cite{SN94}. Discrete Z-symmetries, motivated by the need to improve
the fine-tuning issues \cite{AG97}, can also play a selective
r\^ole. In particular, the spurion gauge singlet superfield $S$ can be
present or not in non-renormalizable operators in the K\"ahler or
superpotential depending on its $R-$ and $Z-$ charges
attributions. Such operators, provided that they do not destabilize
the mass hierarchies \cite{Abel96}, \cite{JBPR}, would lead to
tree-level suppressions of the form $(\langle S \rangle/m_{\rm
Pl})^{2n} \sim (M_X/m_{\rm Pl})^{2n}$. However the decay width is now
too suppressed to yield reheating before BBN except in a very narrow
region centered on $m_{3/2}\sim 1\,$MeV and $M_X\sim 10^{10}\,$GeV.

\begin{figure*}
  \centering
  \includegraphics[width=0.9\textwidth,clip=true]{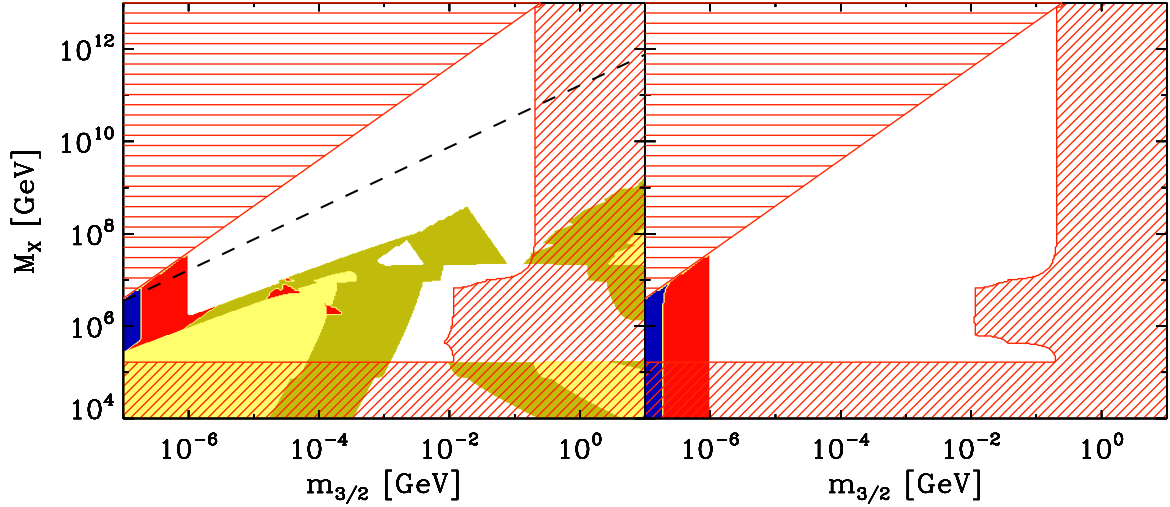}
  \caption[...]{Contours of $\Omega_{3/2}$ in the plane $M_X-m_{3/2}$
  for one pair of messengers sitting in
  $\mathbf{16}+\mathbf{\overline{16}}$ representations of $SO(10)$;
  the lightest messenger $X$ is a singlet under $SU(3)\times
  SU(2)\times U(1)$. Its loop-suppressed annihilation cross-section
  scales as $(\alpha/4\pi)^4/M_X^2$, and it decays into sparticles
  through non-renormalizable operators with width $\Gamma \sim
  10^{-3}M_X^3/m_{\rm Pl}^2$. Color shading is as in
  Figs.~\ref{fig:fast-T}, ~\ref{fig:FY}. See text for details.\label{fig:SO10a}}
\end{figure*}

  In contrast, a larger number of seemingly generic solutions to the
gravitino problem/gravitino dark matter may be found if one considers
$SO(10)$ grand unification and $ M_S > M_X$. 
This case is studied analytically in a companion 
paper~\cite{LMJ05} for $M_X\sim 10^6\,$GeV. In Section~\ref{sec:2}, it
has been shown that the amount of entropy produced depends
directly on the relic abundance of the lightest messenger, which in
turn depends directly on its nature. In $SO(10)$, 
the lightest messenger is a $SU(3)\times SU(2) \times U(1)$
singlet ($\tilde\nu_R-$like), hence its annihilation cross-section is
suppressed: it may either annihilate through one-loop diagrams or at
tree level into goldstinos, and (at tree level) through suppressed
diagrams of GUT mass bosons exchange. One may estimate the relic
abundance of the lightest messenger in this case by using the
dimensional estimate $\langle\sigma_{XX\to \ldots}v\rangle \sim 
(\alpha/4\pi)^4/M_X^2$ for one-loop diagrams and the annihilation
cross-section into goldstinos computed in Section~II.B (see also 
Ref.~\cite{LMJ05}). One finds that the relic abundance is larger than 
for the $SU(5)$ case, hence the amount of entropy production is
expected to be correspondingly larger.  One then finds that
a generic gravitational decay width $\Gamma \sim 10^{-3} M_X^3/m_{\rm
Pl}^2$ can lead to natural
solutions for gravitino dark matter in a significant part of parameter
space, as shown in Fig.~\ref{fig:SO10a}.

  This solution is attractive for several reasons. As seen in the left
panel of Figs.~\ref{fig:SO10a}, gravitino cold dark matter with the
right relic abundance occurs in a rather large part of parameter
space, hence it appears ``natural'' in this sense. Moreover the decay
width to sparticles assumed $\Gamma\sim 10^{-3}M_X^3/m_{\rm Pl}^2$ is
generic as it is predicted by most non-renormalizable operators that
violate messenger number by one unit. In this region of parameter
space, the factor $k\equiv F_S/F \approx 10^{-5}-10^{-2}$, hence
gravitino dark matter would be obtained for indirect gauge mediated
scenarios; in Ref.~\cite{LMJ05} it is argued that this case can be
naturally incorporated in the simplest indirect GMSB
scenarios~\cite{GMSB}. Furthermore, the solution for gravitino dark
matter occurs in the predictive region in which multi-goldstino
production satisfies the unitarity bound, unlike the solutions seen
hitherto for $SU(5)$. Finally, Fig.~\ref{fig:SO10a} assumes, as the
previous figures, that the NLSP is a bino; for a stau NLSP, the BBN
constraints at large $m_{3/2}$ would be relaxed, and this would
enlarge in turn the space of solutions for $\Omega_{3/2}$.

  One can show~\cite{LMJ05} that the amount of entropy production does
not hinder successful leptogenesis at high reheating temperatures;
this is all the more interesting as leptogenesis scenarios typically
operate in models of $SO(10)$ grand unification rather than $SU(5)$.
Strictly speaking, the above scenario requires the spurion field to be
heavier than the lightest messenger. This issue is however
model-dependent as was briefly discussed in section II.B. For
completeness, we illustrate in the left panel of Fig.~\ref{fig:SO10a}
the opposite configuration where the lightest messenger annihilation
into a pair of spurion fields is controlled by
Eq.(\ref{sigXXSS}). This annihilation leads to a too low messenger
relic density for the entropy dilution mechansim to work.

Finally we note that the results obtained in the present study remain valid when
$R-$parity is violated. In effect, in this case the gravitino lifetime
is $\tau_{3/2}\sim 10^{20}\,{\rm sec}\,(m_{3/2}/1\,{\rm GeV})^{-3}$
for trilinear $R-$parity violating terms~\cite{MC02} or
$\tau_{3/2}\sim 10^{27}\,{\rm sec}\,(m_{3/2}/1\,{\rm GeV})^{-3}$ for
bilinear $R-$parity violating terms~\cite{TY00}. Hence the gravitino
is sufficiently long-lived that it can be considered as stable dark
matter with respect to the formation of large-scale structure. If the
gravitino lifetime $\gtrsim 10^{27}{\rm sec}$ one also finds that
distortions of the diffuse backgrounds due to gravitino decay are
evading observational constraints.  For trilinear $R-$parity violating
terms, this requires $m_{3/2}\lesssim 10\,$MeV, while for bilinear
terms, $m_{3/2}\lesssim 1\,$GeV is sufficient. With regards to the
NLSP, its decay can proceed into visible sector particles on a short
timescale and BBN constraints can be evaded, albeit they are replaced
with constraints on diffuse background distortions. Hence the plots in
parameter space would look similar to what has been found above.

\section{Conclusions}\label{sec:6}

We have presented an exploratory though detailed investigation of
relic LSP gravitino abundances in scenarios of gauge mediated
supersymmetry breaking (GMSB).  This study focuses on the possibility
of gravitino dark matter and on solving the light gravitino
overproduction problem for reheating temperatures after inflation that
are "arbritrarily" high.  GMSB scenarios contain intermediate mass
scale $10^4\,{\rm GeV}\simle M_X\simle 10^{12}\,$GeV messenger fields
which by virtue of their gauge interactions are easily produced in the
primordial plasma. Cosmology requires these particles to subsequently
decay as they would otherwise overclose the Universe (except for a
lightest messenger with $M_X\sim 10-30\,$TeV).  Flavor-changing
neutral currents impose somewhat restrictive limits on messenger
number violating Yukawa interactions, possibly arguing for such
messenger number violation to be rather weak.  If so, the delayed
decay of messengers may subsequently dilute any pre-existing gravitino
abundances in accord with cosmological constraints.

We have thus investigated a fairly complete set of renormalizable and
non-renormalizable messenger number violating operators within supersymmetric
unification in $SU(5)$ (as well as some within $SO(10)$) and their
impact on relic gravitino abundances. Results are shown for a variety
of operators and imposing relevant constraints on NLSP decay and
messenger decay from BBN, as well as constraints on the "warmness" of
gravitino dark matter from the required sucessful formation of
large-scale structure.  With respect to prior, less detailed,
studies~\cite{DGP96,BM03,FY02,FY02b,FIY04,TY00}, we have uncovered a
number of significant changes, notably the importance of
messenger-messenger annihilation into two goldstinos in part of the
$M_X$ - $m_{3/2}$ parameter space, which modifies the messenger pre-decay
freeze-out abundances in $SU(5)$ and $SO(10)$ grand unification.

In general, we have found that  gravitino dark matter in $SU(5)$
grand unification in scenarios with high post-inflationary reheating 
temperatures $T_{\rm RH}$  is only possible for a few specific messenger-matter 
couplings. Furthermore we have shown that these models predict
gravitino dark matter in regions of parameter space in which
messengers annihilation to goldstinos violates  unitarity
 unless one makes specific assumptions on the mass spectrum of
GMSB models, and in particular, that the spurion $S$ be much lighter than
the lightest messenger. 

  In contrast, in $SO(10)$ grand unification gravitino dark matter may
be obtained for a variety of generic operators and in the predictive region of
parameter space where multi-goldstino production is under control, 
as long as renormalizable messenger number violating interactions 
in the superpotential are absent~\cite{LMJ05}.  We
thus believe that gravitino dark matter in GMSB scenarios is a viable
alternative to neutralino (and gravitino) dark matter in supergravity scenarios, and as such deserves further detailed study.

\end{document}